# Competitive Instability in Judo: The Hidden Mechanism within an AI-Driven Non-Linear Dynamics Framework

**Author** Attilio Sacripanti ***IJF Academy***

**Acknowledgments** The author gratefully acknowledges Romeo Fabi for his friendly and valuable assistance.

## Abstract.

This study presents a unified nonlinear dynamical framework for understanding, modelling, and teaching competitive judo. The Tori–Uke dyad is formalised as a constrained multibody system whose behaviour emerges from symmetry breaking, coupling dynamics, and transitions between attractor basins. Two fundamental instability archetypes—rotational collapse (Uchi-mata type) and gravitational-lever collapse (Seoi-otoshi, suwari version)—are identified as the core pathways through which all throwing techniques evolve.
A Functional Instability Index I(t) is introduced as a dimensionless order parameter. Derived from local bifurcation analysis, it integrates geometric, dynamic, and coupling-related variables through a multiplicative nonlinear structure, enabling the quantification of critical transitions such as Kuzushi, Tsukuri, and Kake. Fractional Brownian Motion models the global displacement of the dyad, where the local Hölder exponent encodes the informational structure of the interaction.
An AI-based pipeline extracts instability signatures from high-frequency competition video, providing objective measures such as finite-time Lyapunov exponents, attractor topology, and coupling stiffness. Building on these principles, a three-level teaching framework is proposed, shifting judo pedagogy from a technique-centred to an instability-centred approach.
This study establishes the first theoretical foundations for a predictive science of judo performance and outlines future directions for empirical validation, athlete monitoring, injury-risk modelling, cross-sport applications qnd judo as neurological rehabilitative tool for Parkinson diseases.

## . Introduction

The formalization of the interaction within combat sports under a rigorous physical framework is not entirely new. Early attempts to define the Tori-Uke dyad through the lenses of classical rational mechanics and structural symmetries successfully mapped the macroscopic forces at play: such as gravitational loads, push/pull forces, and friction boundaries—by introducing interaction gradient potentials borrowed from molecular physics, such as Morse and Lennard-Jones archetypes (Sacripanti, 1997). However, while that foundational hybrid deterministic-statistical model accurately described the boundary conditions and macroscopic constraints of the contest, it couldn't fully account for the emergent properties, continuous symmetry breaking, and non-deterministic tactical exploration typical of high-level judo.
The present work represents the natural evolution of that framework, shifting the paradigm from classical interaction potentials to a fully non-linear dynamical systems approach driven by stochastic structures and phase transitions.
Judo competition is not a sequence of isolated techniques but a nonlinear dynamical system (Stergiou & Decker, 2011; Strogatz, 2024) in which two athletes form an evolving interpersonal synergy (Krabben, 2019). Contact through Kumi-kata creates a shared configuration space where the movement of one athlete immediately reshapes the possibilities of the other (Kelso, 1982-2012; Sacripanti, 2011). Earlier work by the author on Kumi-kata formalised grips as biomechanical and informational constraints, anticipating recent findings from AI-based posture analysis (Kato & Yamagiwa, 2022). Ecological-dynamics studies confirm that attack decisions emerge from transitions in interpersonal coordination rather than from pre-programmed actions (Lopes, 2024). The dyad behaves as a single system whose stability is highly sensitive to micro-perturbations, timing, and grip-induced constraints (Preatoni et al., 2010, 2012; Krabben, 2019). To formalise this interaction, the dyad is modelled as a constrained multibody system governed by physical constraints and tactical generalised forces modelled via operational-space impedance control (Banerjee, 2016; Wang, 2023).
"Its global displacement on the tatami follows a stochastic process described by Fractional Brownian Motion, where the local Hölder exponent $h_t$ encodes the informational structure of the fight (Sacripanti, 2016, 2021). Within this nonlinear landscape, instability emerges as the central organising principle of judo: not an error, but the mechanism that drives the transition from symmetry to collapse.
Building on this perspective, the present study introduces two fundamental instability archetypes,vertical rotational instability (Uchi-mata model) and gravitational collapse instability (Seoi-otoshi, suwari version) treated as dynamic abstractions rather than techniques. These archetypes define a continuum onto which all throwing actions can be mapped (Sacripanti, 1987, 2021), each characterised by specific dynamical signatures such as Lyapunov divergence, attractor formation, and nonlinear amplification (Winter, 2023; Strogatz, 2024). This leads to a new classification of judo, grounded not in static mechanical categories but in the instability pathways that govern the evolution of the dyad.
To quantify the evolving loss of stability, we introduce the Functional Instability Index $I(t)$, a dimensionless measure integrating geometric, dynamic, and coupling-related variables: COM displacement, finite-time Lyapunov exponents, asymmetry, rotational torque, downward acceleration, and grip stiffness. This index provides the formal order parameter that Krabben et al. (2019) identified as missing in combat-sport theory.
Artificial intelligence plays a central role in operationalising this framework. Modern pose-estimation and machine-learning methods can detect micro-instabilities, quantify symmetry breaking, and identify hidden patterns in competition videos (Kato & Yamagiwa, 2022; Winter et al., 2024; Dindorf, 2024). These tools allow the extraction of instability signatures imperceptible to the human eye, enabling new analytical, classificatory, and didactic applications.
Finally, this work proposes a three-level pedagogical model derived from nonlinear dynamics, aimed at teaching athletes to perceive instability, navigate meta-stable regions, and exploit critical transitions. As a foundational theoretical contribution, this study does not present experimental measurements; rather, it establishes the conceptual and mathematical framework upon which future empirical studies may be built. It represents a coherent attempt to articulate a complex-systems perspective on judo, grounded in decades of theoretical development and supported by appropriate nonlinear and AI-based methods.
The theoretical Instability index was also utilised in a translational applications, as rehabilitative tool for Parkinson's disease (Sacripanti A. 2022). Its effectiveness, quantitatively demonstrated in a recent IJF Academy study, (Sacripanti A. Fabi R., Galea E. 2025) illustrates the broader potential of a systemic approach to judo for motor learning, neuroplasticity, and functional recovery. From this quantitative single case study, in the next September will start a new pilot project in San Marino Republic, named Judo to RISE on Parkinson rehabilitation by Judo.

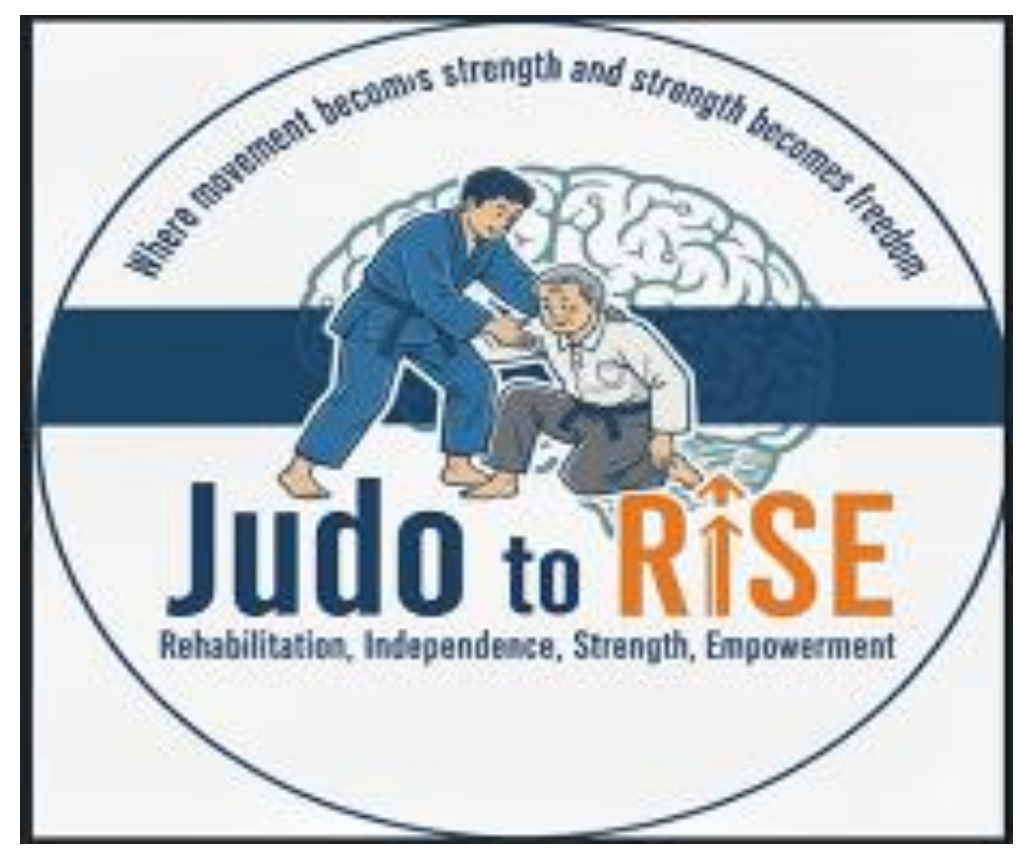

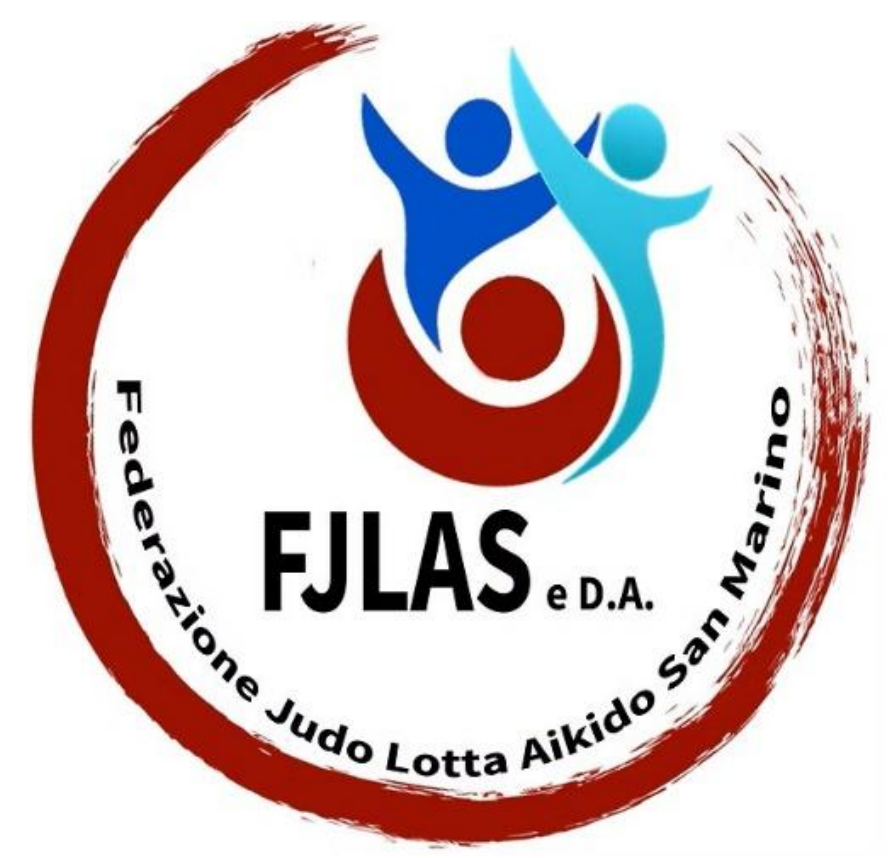

## 2. Multibody Formalism and Tactical Impedance Control in the Dyad

During the transition phase of a throwing action, the two athletes can no longer be considered as independent systems. Instead, they evolve into a single, coupled configuration, the dyad, whose behaviour must be described through the dynamics of a constrained multibody system. To formalise this interaction during the Tsukuri and Kake phases, we introduce the following equation of motion (Wang, 2023; Banerjee, 2016; see Appendix B):

$J_{\text{grip}}(q) = \frac{\partial x_{\text{grip}}}{\partial q}$where:

- $q$ is the vector of generalised coordinates,
- $M(q)$ is the mass matrix,
- $M(q)\ddot{q} + N(q,\dot{q}) + G(q)$ represents the intrinsic dynamics of the body segments, including nonlinear Coriolis, centrifugal, and gravitational forces,
- $J_q^T$ is the Jacobian matrix of constraints (Trench, 2013).

Physical constraints

$\mu_{\max}(t) = \max_i \mathrm{Re}\left(\lambda_{\mathrm{i}}\big(A(t)\big)\right)$ represents the classical Lagrange multipliers associated with physical constraints: foot–tatami contact (unilateral constraint), grips (bilateral constraints with elastic and dissipative components), elastic collisions between body segments.

Virtual tactical constraints

A key innovation of this model is the introduction of $Q_{tact}$, the vector of generalized tactical control forces (operational space impedance formulation). To avoid the mathematical singularity associated with rigid virtual constraints (Lagrange multipliers), the tactical action of Tori through Kumi-kata is modelled as a time-varying nonlinear viscoelastic system in operational space:

$$Q_{\text{tact}}(q,\ddot{\text{t}},t) = J_{\text{grip}}^T\left[F_{\text{act}}(t) - K_{\text{grip}}(t)e_{\text{x}}(q,t) - D_{\text{grip}}(t)\dot{e_{\text{x}}}(q,\ddot{\text{t}},t)\right]$$

where:

$J_{grip}(q) = \partial x_{grip}/\partial q$ is the kinematic Jacobian at the grip contact points;

$e_x(q,t) = x_{Uke}(q) - x_{Tori}(q) - d_{des}(t)$ is the Cartesian position error relative to the desired inter-segment distance $d_{des}(t)$, encoding Ma-ai management;

$K_{grip}(t) \in \mathbb{R}^3 \times ^3$ and $D_{grip}(t) \in \mathbb{R}^3 \times ^3$ are the active stiffness and damping matrices dynamically modulated by Tori;

$F_{act}(t)$ is the vector of active directional forces (instantaneous push/pull actions).

This formulation eliminates the mathematical singularity of rigid virtual constraints and models Kumi-kata as what it physically is: a variable-impedance dynamic coupling (Krabben, 2019). Variations in grip impedance modify the system's stability landscape, driving the emergence of functional instability.

Muscular input

$Q_{\text{musc}}$ represents the internal metabolic forces transformed into external mechanical torques applied by Tori at the moment of Kake.

### 2.1 Global Motion as Fractional Brownian Motion

Once the internal dynamics of the dyad are defined by Eq. (1), its global displacement on the tatami can be modelled as a stochastic process governed by Fractional Brownian Motion (fBm) (Sacripanti, 1987, 2012, 2021). This framework captures the high-entropy exploratory behaviour typical of the Kumi-kata phase (Sacripanti, 2011). The Eq. (5) models the stochastic energy flow: when $h_t < 0.5$ (anti-persistent, $\Gamma(h_t) < 0$), energy is absorbed by Uke rather than directed by Tori, making $\lambda_{\text{tactics}} \approx 0$Uke retains too many degrees of freedom and active defence prevails, making $Q_{\text{musc}}$ insufficient to overcome inertia.

### 2.2 $Q_{\mathrm{musc}}$ as a Control Parameter and Phase Transition Trigger

From a nonlinear-dynamics perspective, $Q_{\mathrm{musc}}$ acts as the control parameter of the dyadic system (Ladevèze & Pelle, 2004; Amato, 2006). While the dyad moves in a high-entropy state during Kumi-kata (modelled via fBm), the strategic application of $Q_{\mathrm{musc}}$ triggers a phase transition. For Tori, this transition corresponds to the saturation of Uke's degrees of freedom: Tori does not merely apply force, but compresses Uke's available phase space.

As a result: the configurational entropy of the dyad collapses (Vervoorts et al., 2021), the stochastic tactical search becomes a deterministic projection toward the functional attractor (Ippon), and the global stochastic motion is transformed into a deterministic trajectory ending in a lever-based or couple-based mechanical tool (Sacripanti, 1987, 2021).

These multibody equations describe the structural and constraint-based organisation of the dyad. However, the emergence of effective throwing actions cannot be understood through mechanical structure alone. Once physical constraints and impedance-controlled tactical forces shape the configuration space, the dyad evolves according to nonlinear dynamics in which symmetry breaking, divergence, and attractor formation govern the transition from stability to collapse. This motivates the shift to the dynamical analysis developed in Section 3.

## 3. Nonlinear Dynamics of the Dyad

### 3.1 Symmetry, Asymmetry, and Symmetry Breaking

Understanding judo interaction requires distinguishing two fundamental forms of symmetry breaking that underlie effective Kuzushi in competition (Sacripanti, 2014, 2021). Stability depends on the base of support and the position of the centre of mass (COM). We therefore differentiate: static symmetry breaking (postural asymmetry) and dynamic symmetry breaking (interactional asymmetry).

#### A) Static symmetry breaking (postural asymmetry)

This occurs when Uke's configuration is inherently unbalanced before any dynamic action. A rotated torso, lateral lean, or uneven stance shifts the COM and creates a stability gradient. Uke may appear stable on one side but becomes globally less mobile. This reduction in mobility provides Tori with a predefined attack direction. The system's symmetry is broken by geometry alone, allowing Tori to exploit a structural weakness without initially injecting high kinetic energy (Sacripanti, 2014, 2021).

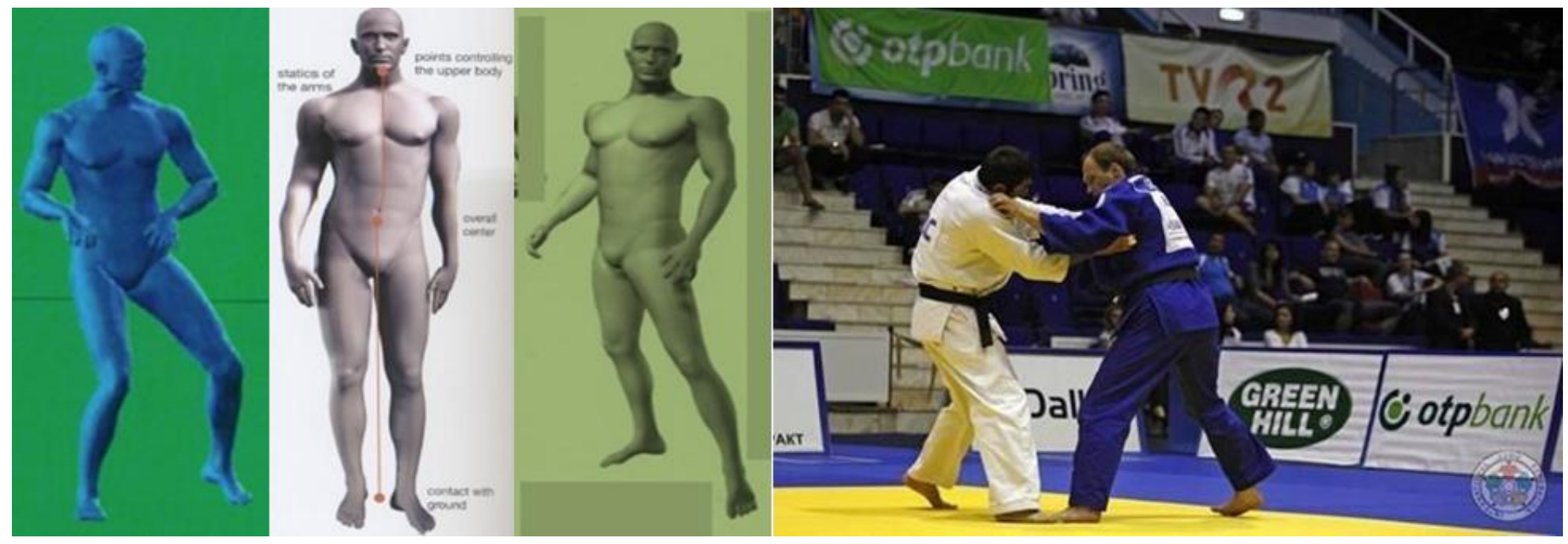

***Fig. 1 – Static postural symmetry breaking: Blue targeting White right side***

#### B) Dynamic symmetry breaking (interactional asymmetry)

Dynamic symmetry breaking emerges from the coupling of movement and Tori's kinetic input. The dyad transitions from a reciprocal equilibrium to a chaotic regime characterised by a positive Lyapunov exponent (Winter et al., 2023). Kato & Yamagiwa (2022) demonstrated that Kumi-kata acts not only as a mechanical constraint but also as an informational constraint: the configuration of arms, torso, and contact predicts the direction of instability and the family of emerging techniques. Small perturbations, rhythm changes, or entries (Tsukuri) can drive the system from reciprocal exchange to divergent instability (Kuznetsov, 2023).

Identifying and exploiting both forms of symmetry breaking is central to effective Kuzushi (Sacripanti, 2021). Once symmetry is broken, the dyad no longer evolves in a neutral configuration space. The interaction becomes progressively channelled toward specific coordination tendencies, or attractors, which define the functional landscape of instability.

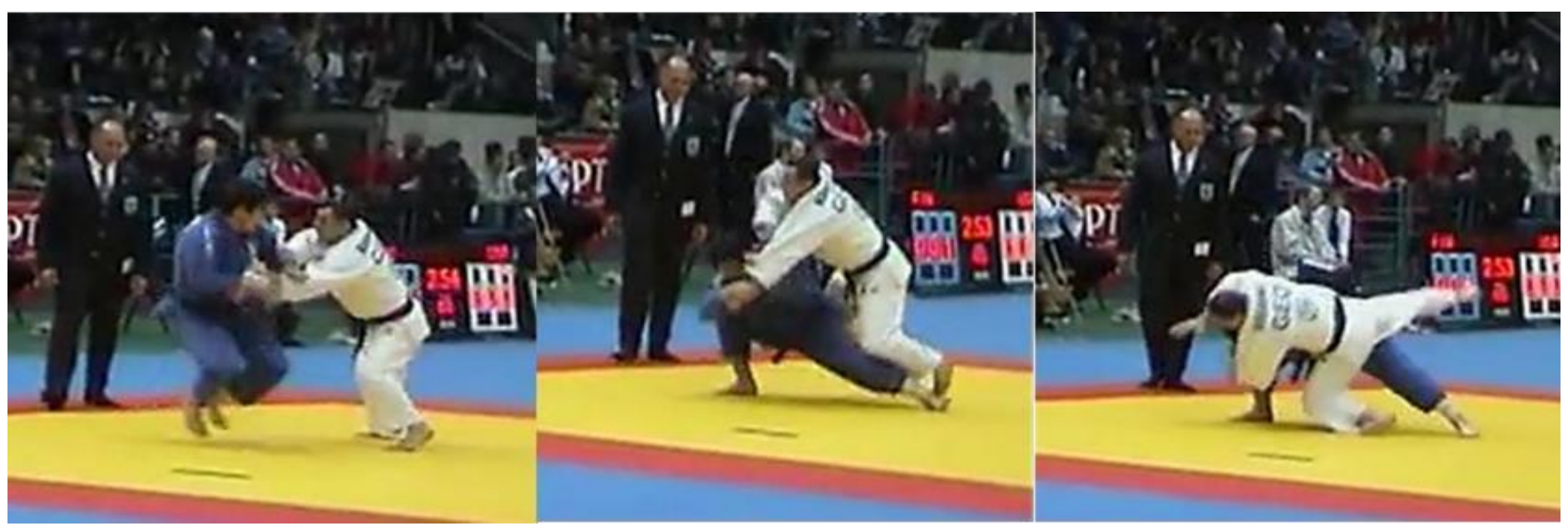

***Fig. 2 – Dynamic symmetry breaking, interactional asymmetry action in competition***

*Practical application: identifying the target*
Coaches should train athletes to "read" static asymmetry. When Uke presents a rotated torso, certain degrees of freedom become effectively locked. AI can quantify this by measuring COM deviation relative to the base of support and identifying the side where defensive response is slowest.

## 3.2 Attractors and Functional Instability: Instability as the Core of Judo Dynamics

A technique should not be viewed as an isolated motor pattern but as a trajectory in the configuration space of the dyad (Riley & Turvey, 2002). Lopes (2024) showed that the fight evolves through stable, meta-stable, and unstable states, and that attack decisions correspond to transitions between coordination patterns. This supports the view that instability is an emergent property of the dyad and represents the dynamic core of competitive judo. In dynamical systems, an attractor is a region of phase space toward which the system tends to evolve (Strogatz, 2024; Stergiou & Decker, 2011). In judo, attractors correspond to functional tendencies such as preferred imbalance directions, characteristic rhythms, or rotational pathways that facilitate instability. Tori's task is not simply to "apply a throw," but to drive the dyad toward a functional attractor (Winter et al., 2024; Yearby et al., 2024) that reduces Uke's ability to recover balance. A throw succeeds when the system enters an unstable basin where collapse becomes inevitable, like Tai Atari, (Strogatz, 2024), after which the mechanical tool (lever or couple) finalises the projection (Sacripanti, 1987, 2021).

- A. Instability as an emergent property of the dyad: Tori and Uke form a coupled system in which each micro-action instantly modifies the other's capabilities. Krabben et al. (2019) describe this as an "interpersonal synergy." Instability arises when this synergy is disrupted.
- B. Instability as manipulation of interpersonal synergy: Instability emerges from symmetry breaking, local divergence (Lyapunov), COM variations, changes in coupling stiffness Kg(t), and transitions between attractor basins. Skilled performance requires operating in meta-stable regions (Krabben, 2019). The Instability Index I(t) formalises this concept.
- C. Instability as a critical transition (bifurcation): Every technique crosses a critical threshold where the dyad transitions from stability to collapse. Small changes in constraints can cause sudden transitions (Krabben, 2019). This is the essence of nonlinear amplification and explains why Tori must guide the dyad toward a functional attractor (Winter, 2023).

### 3.3 Functional Instability Index $I(t)$

To quantify evolving stability, we introduce the Instability Index $I(t)$, a dimensionless order parameter I(t) ∈ [0,1] derived from local bifurcation analysis of the multibody system (Correction 3). Rather than an ad hoc additive heuristic, I(t) integrates dynamical stability (Jacobian eigenvalues), geometric stability (support polygon distance), and inter-athlete coupling (grip impedance) in a multiplicatively nonlinear structure. See Appendix B for the full derivation. Grip stiffness $K_g(t)$. Kato & Yamagiwa (2022) showed that Kumi-kata posture contains predictive signals about future behaviour. Lopes (2024) demonstrated that transitions can be quantified through collective variables such as relative angular velocity and running correlation. The Instability Index extends this logic by integrating multiple variables into a single dimensionless measure (see Appendix B). **Definition (normalized form)**

$$\chi(t) = \frac{\mu_{\max}(t)}{\eta(q(t))} \cdot \frac{K_0}{K_{\text{grip}}(t) + K_0}$$

$$\eta(q) = 1 - \frac{\left\| x_{\text{COM}}(q) - x_{\text{BOS,center}} \right\|}{R_{\text{BOS}}(\theta)}$$

$$I(t) = \frac{1}{1 + e^{-k(\chi(t) - \chi_{\text{crit}})}} \quad (2)$$

***Biomechanical rationale***: Geometric contributors ($S$, $A$, $T_{\text{rot}}$) describe spatial evolution toward instability. Dynamic contributors ($\lambda$, $a_z$) quantify divergence rate. Stabilizing contributors ($K_g$) represent resistance to instability through grips

- D. Instability as hidden order: Instability follows attractors, basins, Poincaré maps, FTLE, and the Hurst parameter. Krabben et al. (2019) emphasise the need for order parameters; the archetypes Uchi-mata and Seoi-otoshi provide such structures.
- E. Instability as a tactical tool: Instability is induced, not endured. Eq. (1) (constraints), Eq. (5) (energy flow), and Eq. (2) (Instability Index) formalise this process.
- F. Instability as a measurable variable: The Instability Index is the first quantitative proposal integrating geometry, dynamics, coupling, and acceleration to identify collapse windows and distinguish instability archetypes.
- G. Instability as expert skill: The three-level teaching model trains athletes to perceive instability, navigate meta-stable regions, and act at the singularity point. Krabben et al. (2019) describe this as expert brinkmanship.

#### 3.3.1 Two Fundamental Instability Archetypes

The dyad transitions from stability to collapse through two fundamental instability pathways.
These archetypes are not techniques but dynamic structures that organise the evolution of the Tori–Uke system.

**I) Rotational Archetype (Uchi-mata Type)**

Instability emerges through the progressive formation of a rotational attractor. Tori applies a lateral-vertical couple of forces that generates a dominant torque around an oblique axis. The resulting angular dynamics increase Uke's rotational velocity until recovery becomes impossible. The COM trajectory exhibits a characteristic lateral deviation, and collapse occurs when the projection of the COM exits the support polygon. This archetype corresponds to a smooth, rotationally driven loss of stability.

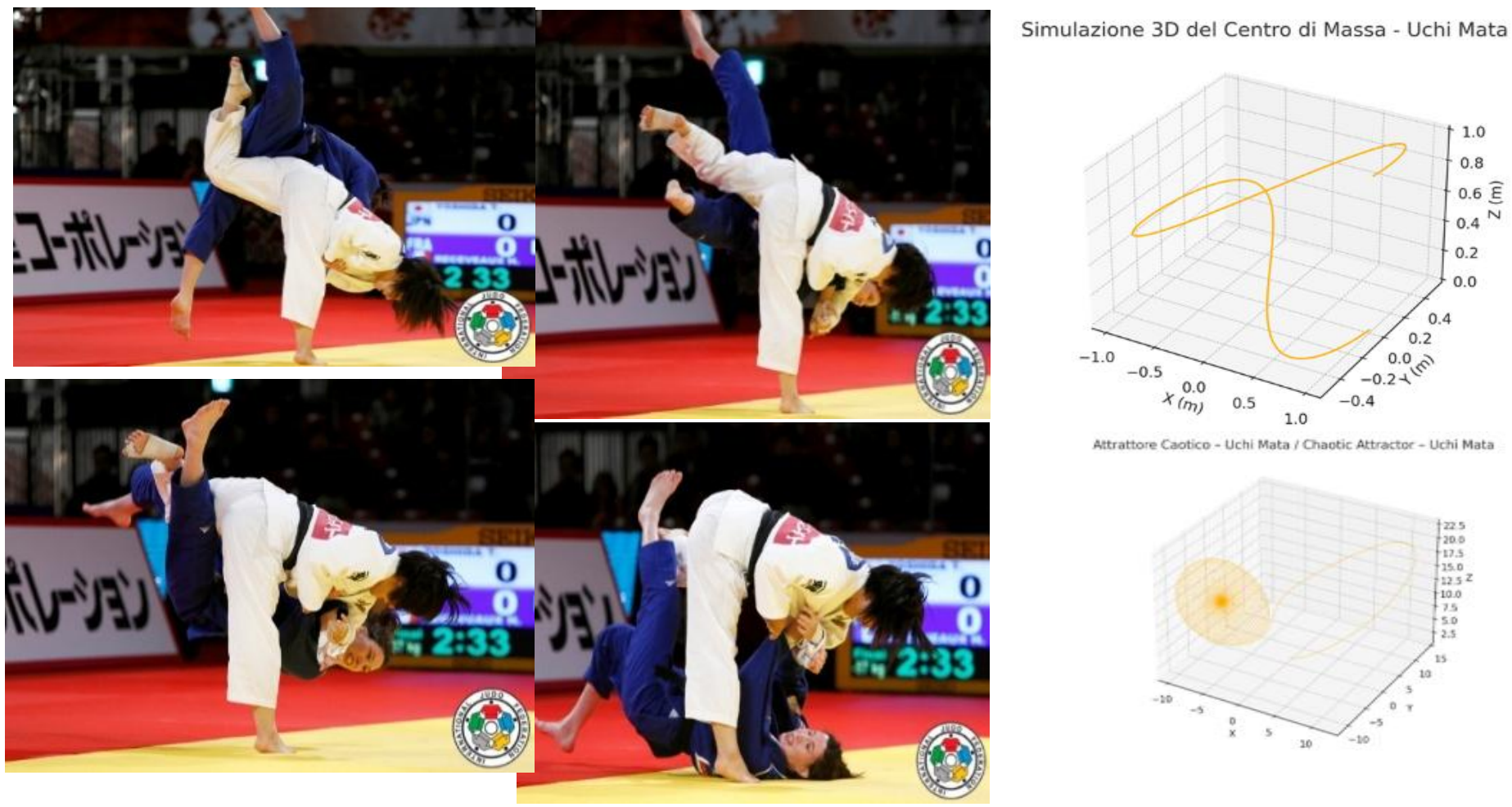

***Fig. 3 Uchi-mata: loss of symmetry creates an instability collapse. On the side, illustrative representations of the displacement of the COM in real space and Uchi-mata's chaotic circular attractor in phase space***

**II) Gravitational-Lever Archetype (Seoi-otoshi, suwari version)**

This archetype combines:

1. a rapid drop of Tori's COM, creating a vertical void beneath Uke, and
2. a lever action applied through a migrating fulcrum.

The effective fulcrum is not fixed. It initially lies at the feet–tatami contact, where friction provides transient resistance. Once friction is overcome, the fulcrum migrates to the contact between Uke's shins and Tori's crouched body, which becomes the main pivot of the lever. The shoulder acts as a strong kinematic constraint but not as a fixed point. Because the fulcrum moves and the applied force has both vertical and lateral components, Uke's COM follows a descending helical trajectory. Collapse is abrupt and irreversible once the COM projection leaves the support polygon. This archetype represents a mixed gravitational- lever instability with characteristic helical fall. A full mathematical treatment of both instability archetypes is provided in Appendix

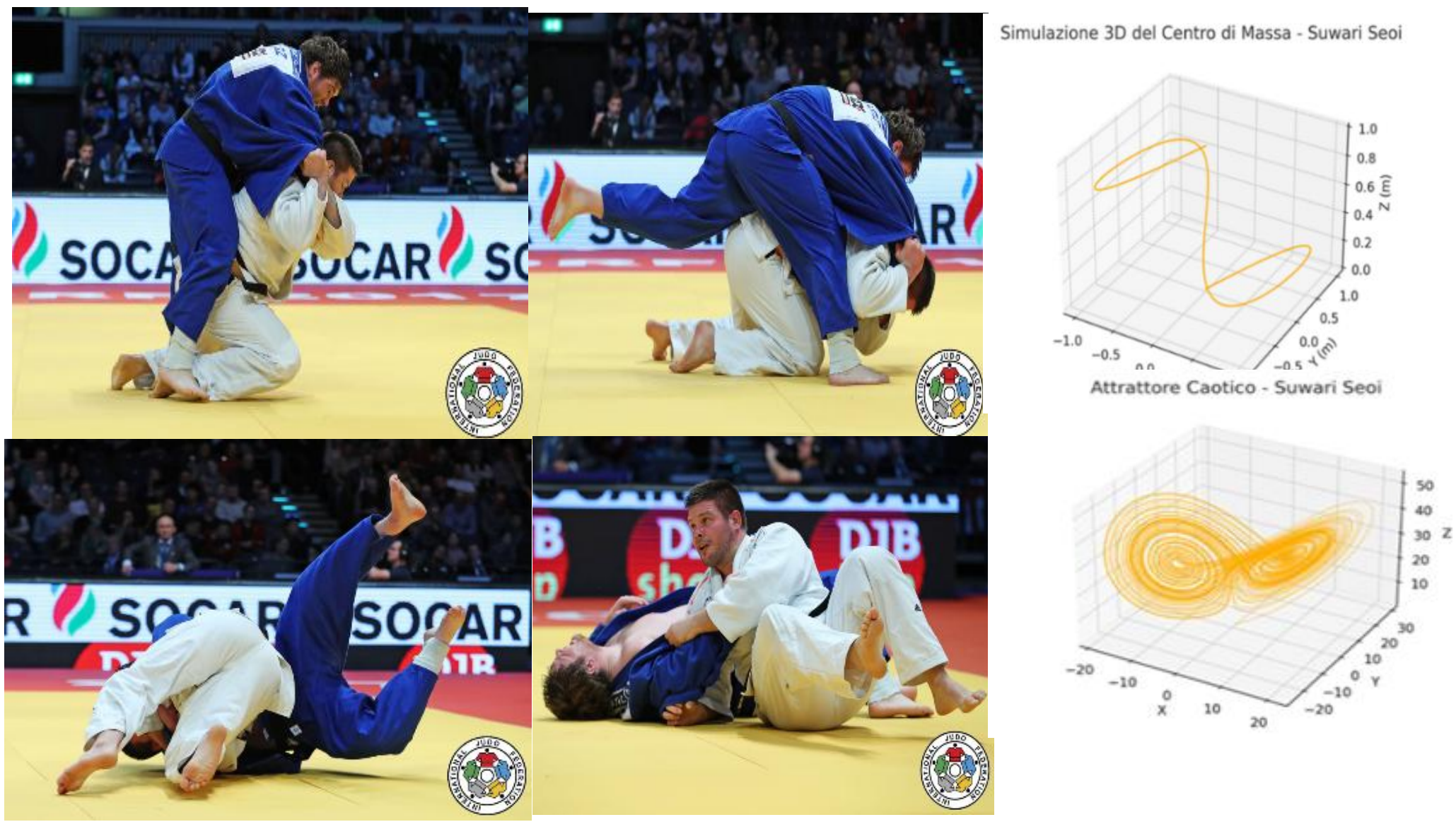

***Fig. 4 – Seoi-otoshi, suwari version: the functional asymmetry required to generate momentum (higher energy and lower stability). On side, illustrative representations of the displacement of the COM in real space and collapse chaotic attractors in Seoi-otoshi indicating motion variability and internal dynamics.***

### *3.4 A New Instability-Based Classification of Judo Throws*

Nonlinear analysis shows that the core of competitive dynamics is not the technique itself but the instability induced within the dyad. Each technique corresponds to a specific instability basin. Tori then applies one of the two operational tools—lever or couple—to finalise the projection. All techniques can be ordered along a continuum between the two archetypes: Uchi-mata and Seoi-otoshi (suwari variation). This classification complements Kano's five principles and Sacripanti's biomechanical classification (Kano,1895, 1986; Sacripanti, 1987, 2021).

If instability archetypes define the qualitative structure of throws, phase transitions describe the precise moment in which the dyad crosses the threshold from stability to collapse. Section 3.5 formalises these transitions using nonlinear tools such as Lyapunov exponents and Poincaré maps.

### 3.5 Phase Transitions, Thresholds, and Nonlinear Amplification

Every technique passes through a critical threshold where the dyad transitions from stability to instability (Riechl, 1992; Kuznetsov, 2023). Small changes in COM, timing, or entry angle can produce nonlinear amplification. Quantitative tools such as Lyapunov exponents and Poincaré maps measure divergence and regime changes (Winter, 2023; Strogatz, 2024).

**Lyapunov exponent**

$$\lambda = \lim_{t \to \infty} \frac{1}{t} \log \frac{\|\delta x(t)\|}{\|\delta x(0)\|} \qquad (3)$$

A positive value indicates rapidly growing instability (Strogatz, 2024). In judo, finite-time Lyapunov exponents (FTLE) are used due to the transient nature of interaction.

**Poincaré map**

$$\Sigma \to \Sigma_x : x_n \mapsto x_{n+1} \qquad (4)$$

Useful for identifying recurring interruptions in rotation or rhythm (Parker & Chua, 1989).

### 3.5.1 Phase Transition in Kumi-kata

Grips transmit both mechanical force and informational signals (Sacripanti, 2011). The energy flow between Tori and Uke can be expressed as:

$$dE(t) = \left(F_{\text{kumi}}(t) \cdot \Delta v_{\text{COM}}(t)\right)dt + \Gamma(h_{\text{t}})P_0\, dt + \sigma_{\text{stoch}}(t)dB_{h_{\text{t}}}(t) \qquad (5)$$

where:

- $F_{\text{kumi}}(t)$: force transmitted through the grip (the "control vector"),
- $\Delta v_{\text{COM}}(t)$: difference in velocity between the centres of mass,
- Local Hölder exponent $h_t \in (0,1)$ at instant t, estimated over a sliding window [t−τ, t]. It quantifies local regularity without requiring differentiation of the Hurst parameter.
- The coupling function $\Gamma(h_t) = \tanh(\kappa(h_t - 0.5))$ modulates the base stochastic power $P_0$ (in Watts): $\Gamma > 0$ if $h_t > 0.5$ (persistent, Tori accumulating dominance); $\Gamma < 0$ if $h_t < 0.5$ (anti-persistent, Uke absorbing energy via rapid micro-corrections),

$$\Gamma(h_{\text{t}}) = \tanh\left(\kappa(h_{\text{t}} - 0.5)\right)$$

- $F_{\text{kumi}}(t)\, \Delta v_{\text{COM}}(t)$: deterministic mechanical power component (grip-mediated)
- stochastic noise amplitude $\sigma_{stoch}(t)$ induced by tactical uncertainty of the study phase (in $W \cdot s^{-1/2}$), driving the local fBm term $dB_{ht}(t)$

***Practical Example***: when the pressure rhythm of the kumi-kata becomes persistent (h>0.5), it is time to attack. If, however, Uke, absorbing the pressure, continues to change direction (h<0.5), it is better to persist with the study to find or produce a symmetry breaking moment.

## 4. Dyadic Displacement and Internal Dynamics

To understand the "hidden order" of judo, the global movement of the Tori–Uke dyad must be decomposed into two coupled scales:

1. the macro-scale displacement of the dyad on the tatami, and
2. the micro-scale internal dynamics between the athletes (Sacripanti, 2021; Wang, 2023).

Traditionally, global displacement has been interpreted as a sequence of random steps. Here, it is modelled as a stochastic process governed by Fractional Brownian Motion (fBm). Recent advances in deep learning and computer vision (Zheng et al., 2024; Rojas-Valverde et al., 2025) allow the detection of subtle symmetry-breaking events within this motion. These AI-driven insights reveal that the transition from Tsukuri to Kake corresponds to a bifurcation point, where the system's stability is intentionally compromised to achieve the throw.

### 4.1 Fractional Brownian Motion

To capture the multiscale structure of judo, global dyadic motion must be separated from micro-scale instability. As shown in previous work (Sacripanti, 1987, 2015, 2021), the movement of the dyad does not follow a linear path, a standard random walk, or classical Brownian motion. Instead, it is better described by Fractional Brownian Motion (fBm).

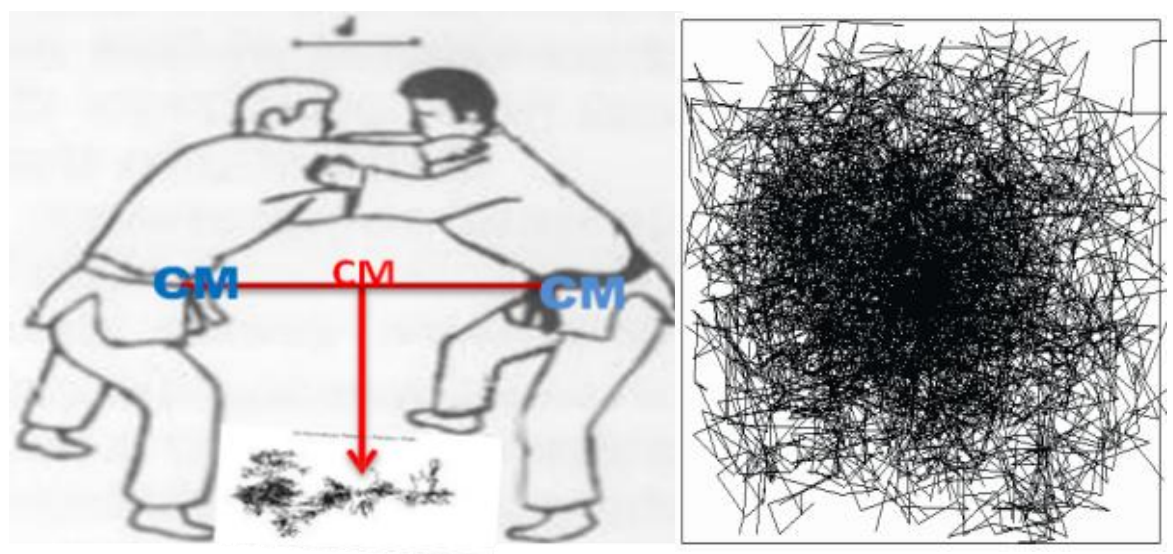

***Fig. 5 – Trajectories on the tatami of the collective COM of a couple of athletes (dyad). Realization for 16 virtual competitions (Sacripanti, 1987, 2015, 2021)***

**Mathematical implication**

The trajectory of the dyad's collective COM exhibits long-range dependence or memory (Meerson et al., 2022). Movement is not a sequence of independent steps but a persistent or anti-persistent process in which past displacements influence future ones.

$$H = \lim_{T \to \infty} \frac{\log \mathrm{Var} \mid X(t+T) - X(t) \mid}{2 \log T} \qquad (6)$$

**Practical takeaway (Hurst exponent)**

Bryce & Sprague (2012) showed that short, non-stationary time series require robust methods to estimate long-range dependence; many popular techniques (e.g., DFA) introduce biases in nonlinear contexts. The Hurst parameter indicates whether dyadic motion tends to: persist in the same direction ($H > 0.5$), reverse direction frequently ($H < 0.5$), or behave as memoryless noise ($H \approx 0.5$).

### 4.1.1 Global Motion Model

Once the internal dynamics of the dyad are defined by Eq. (1), its global displacement is modelled as an fBm-driven stochastic process (Biagini et al., 2008). In this framework, fBm becomes an indicator of strategic dominance: movement on the tatami is not noise but a search for a critical configuration. Correct interpretation requires distinguishing between fGn and fBm signals (Eke et al., 2002). The dyad's COM evolution can be expressed through the fractional stochastic differential equation:

$$dX(t) = \mu(X,t)\,dt + \Sigma\,dB_H(t) \qquad (7)$$

where:
$\mu(X,t)$: deterministic drift (Tori's biomechanical intention),
$dB_H(t)$: increment of fractional Brownian motion governed by $H$.

**Interpretation in judo competition** (Sacripanti, 1987, 2015, 2021):

a. Study phase ($H \approx 0.5$): motion is stochastic; neither athlete controls the dyad.
b. Attack/persistence phase ($H > 0.5$): symmetry is broken; movement becomes directional; this corresponds to Kuzushi leading to Tsukuri.
c. Defence/anti-persistence phase ($H < 0.5$): Uke performs rapid corrective actions to restore balance.

**Practical application**

For coaches, this confirms that the "tactical dance" before a throw is not random but a search for an optimal configuration in both real space and phase space.

### 4.2 Individual Dynamics Beyond Classical Mechanics

While the dyad moves according to fBm, the individual movement of each athlete during a technique cannot be fully explained by classical mechanics. Instead, nonlinear dynamics provide a more accurate approximation.

Limitations of classical models

Classical mechanics treats the athlete as a system of levers and pulleys in a vacuum. In reality, the athlete is an open system, continuously exchanging energy and information with the opponent and interacting with frictional forces.

Nonlinear individual response

Each athlete behaves as a complex system capable of sudden state transitions. The apparent simplicity of global dyadic movement is an emergent property of highly complex, nonlinear musculoskeletal adjustments.

Symmetry and target identification

As noted in Section 3.1, static symmetry breaking creates a non-uniform stability landscape. Nonlinear models reveal how Tori's movement penetrates Uke's weakest stability points rather than overcoming them through brute force.

Practical Application: Training the "Nonlinear Athlete"

Coaches should shift from teaching static positions to training dynamic situations. AI-based tools can provide: early-warning indicators of symmetry breaking, visualisation of entries, objective timing metrics, stability landscape maps, micro-timing windows, corrective drills, comparative athlete profiles.

These outputs can be integrated into a coaching dashboard to support decision-making and targeted training.

## 5. Discussion, Limitations, and Future Work

The combined analysis of global dyadic displacement (fBm) and micro-scale internal dynamics reveals that competitive judo is governed by multiscale interactions linking stochastic exploration to deterministic collapse. These findings provide the conceptual basis for the broader implications discussed in this section, including theoretical coherence, methodological limitations, and future empirical developments.

### 5.1 Summary of Contributions

This foundational theoretical study proposes a unified framework that reframes judo throwing actions as instability processes emerging from the nonlinear interaction of the Tori–Uke dyad. The principal contributions are:

1. Formalisation of the dyad as a constrained multibody system incorporating both physical and tactical (virtual) constraints (Eq. 1).
2. Introduction of two fundamental instability archetypes, vertical rotational collapse (Uchi-mata model) and gravitational collapse (Seoi-otoshi, suwari version), defining a continuum onto which all throwing techniques can be mapped.
3. Definition of the Functional Instability Index $I(t)$ (Eq. 2, revised per Correction 3): a dimensionless order parameter derived analytically from local bifurcation analysis. It integrates dynamical stability ($\mu_{max}$, maximum real Jacobian eigenvalue), geometric stability (η, normalised COM-to-support-polygon distance), and grip impedance ($K_{grip}$) in a multiplicatively nonlinear structure — replacing the previous additive heuristic — integrating geometric, dynamic, and coupling-related variables.
4. Presentation of three level pedagogical model derived from nonlinear dynamics, aimed at teaching athletes to perceive instability, navigate meta stable regions, and exploit critical transitions.
5. Incorporation of the local Hölder exponent h_t $H$ as an indicator of the informational structure of Kumi-kata and its role in the stochastic-to-deterministic phase transition.
6. Specification of an AI pipeline (Appendix A) for extracting instability signatures from competition video, enabling operationalisation of the framework for coaches and analysts.

### 5.2 Limitations

The limitations of this work fall into three categories: foundational, technical, and domain-specific.

#### 5.2.1 Foundational Limitations

Empirical validation. As a foundational theoretical paper, this work does not provide empirical datasets, effect sizes, or statistical tests to validate thresholds such as Lyapunov exponent sign changes, critical values of $I(t)$, or Hurst parameter ranges. These constructs require validation against labelled competition data. The framework should therefore be interpreted as a roadmap for future empirical research, not as a validated model.

Model identifiability. The study introduces dynamical constructs, attractors, basins of attraction, finite-time Lyapunov exponents, without specifying parameter-identification procedures or uncertainty quantification. Estimating Lyapunov exponents from short, noisy time series (typical of judo exchanges lasting a few seconds) is challenging, and their reliability under competition conditions remains untested.

Generalisability. The framework is presented as universal, but its applicability across weight categories, age groups, genders, and tactical cultures has not been evaluated. Dynamical signatures may vary systematically with morphology, strength-to-weight ratios, or national styles.

#### 5.2.2 Technical Limitations

Parameter estimation from video. The AI pipeline (Appendix A) proposes estimating grip stiffness, rotational torque, and finite-time Lyapunov exponents from standard competition video. These estimates rely on pose-estimation algorithms that introduce errors under occlusion, rapid motion, or suboptimal camera angles. The sensitivity of $I(t)$ to such errors has not been quantified.

Weight calibration. Equation (2) contains six coefficients $(\alpha, \beta, \gamma, \delta, \epsilon, \zeta)$ plus a normalisation constant. These must be calibrated via supervised learning on expert-labelled sequences, but the paper does not provide data on inter-rater reliability or on the stability of learned weights across datasets or competition contexts.

Computational requirements. The proposed pipeline, high-frequency video capture (120-240 fps), 3D pose estimation, multibody dynamics, nonlinear time-series analysis, is computationally intensive. Real-time implementation for in-competition feedback remains technically challenging.

#### 5.2.3 Domain-Specific Limitations

Intrinsic complexity of judo competition. As noted in the author's 2015 *Judo Optimization* presentation (Toledo), competitive judo is a nonlinear system with fractal trajectories and continuously evolving open skills. Coaches often cannot plan optimisation in real time. Figure 6 illustrates rapid transitions between standing (Nage-waza) and ground fighting (Ne-waza), which only high-frequency AI analysis (240-300 fps) can resolve.Transitional and ground phases. The framework focuses primarily on standing throwing actions. Ne-waza dynamics—where the objective shifts from creating instability to creating maximum stability, are only briefly mentioned. A complete dynamical theory must address both regimes and their transitions.

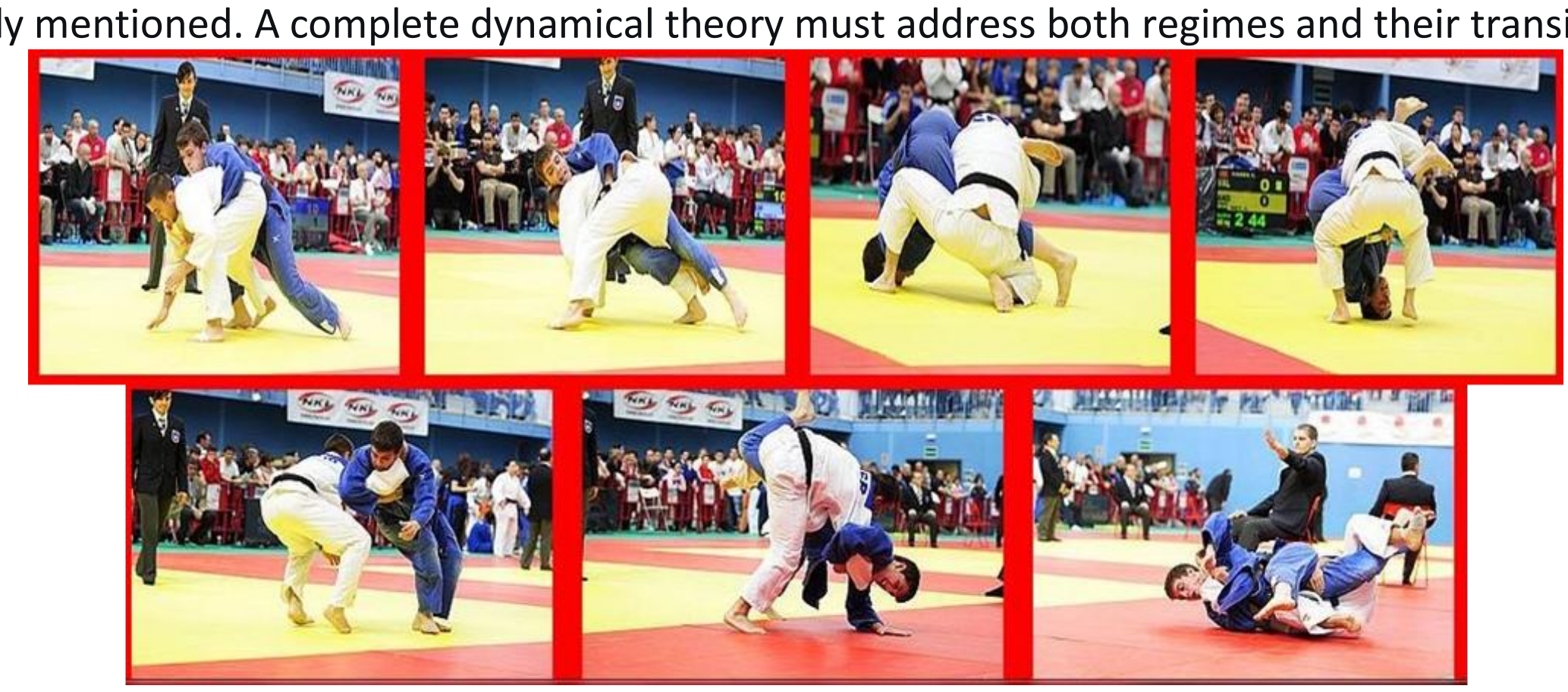

***Fig. 6 – Ne-waza and Nage-waza continuous transition in competition***

Environmental and contextual factors. Competition outcomes depend on factors beyond dyadic interaction: referee decisions, mat-edge constraints, fatigue, psychological pressure, and strategic fouling. These are not captured by the current model. For a practical example of applying the Instability Index to a match situation, see Appendix C.

## 5.3 Future Work

The limitations above define a structured research programme for translating this theoretical framework into validated, practical tools.

Empirical validation studies. The immediate priority is applying the AI pipeline to large, diverse competition datasets—ideally from World Judo Tour events—and validating $I(t)$ against expert-labelled events (Kuzushi onset, Tsukuri completion, Kake initiation). Sensitivity analyses should include weight classes, rule changes, recording conditions, and pose-estimation algorithms.

Parameter estimation and uncertainty quantification. Future work should develop robust methods for estimating dynamical quantities from short, noisy time series, including bootstrap confidence intervals for Lyapunov exponents and Bayesian calibration of $I(t)$. Identifiability of attractor structures should be tested using surrogate data.

Longitudinal athlete monitoring. The framework may support long-term monitoring of elite athletes, identifying the emergence of Tokui-waza attractors, detecting early performance degradation, or quantifying injury effects on dynamical signatures.

Expansion to Ne-waza. The dynamical principles established for throwing techniques should be extended to ground fighting, where stability and instability play complementary roles.  Probably in Ne-Waza the functional Instability index could be adapted  or even hypothetically inverted , because in Ne-Waza Tori, into the dyad, try to obtain the maximum stability: as in Osae-Komi. The transition between Nage-waza and Ne-waza (Figure 6) is particularly rich for nonlinear analysis.

Injury prevention. The relationship between instability signatures (e.g., large or rapidly growing Lyapunov exponents during Sutemi-waza or counter-throws) and injury risk should be investigated.

A validated injury-risk model could provide real-time warnings to referees or medical staff.

Referee support and scoring objectivity. Objective stability metrics derived from $I(t)$ could complement current scoring criteria, especially in borderline cases between Waza-ari and Ippon.

Transfer to other combat sports. The framework's applicability to wrestling, Brazilian Jiu-Jitsu, sambo, and other grappling sports should be evaluated, adapting instability archetypes and AI tools to sport-specific constraints.

Ethical and practical considerations. Future work must address informed consent, data privacy, and potential misuse of performance analytics. Transparency, reproducibility, and open-science practices should guide the transition from theory to application.

# 6. Methods

This study employs a multi-layered analytical approach combining nonlinear dynamics, multibody formalism, stochastic modelling, and artificial intelligence to construct a unified theoretical framework for judo competition. No empirical data were collected or analysed; the methods described here concern the conceptual and mathematical development of the framework. The AI pipeline summarised below is presented in full detail in Appendix A and is included here to contextualise its role within the overall methodology.

## 6.1 Modelling Framework

The core of the framework is the formalisation of the Tori–Uke dyad as a constrained multibody system (Section 2, Eq. 1). The Lagrangian formulation (Appendix B.1, B.3) defines the system's kinetic and potential energy, while constraints are imposed through Lagrange multipliers:

- **physical constraints:** foot–tatami contact, grip coupling, joint limits;
- **tactical impedance control forces Q_tact:** space closure, degree-of-freedom saturation (Makikomi), directional forcing (Tai Atari).

Internal muscular torques $Q_{\mathrm{musc}}$ act as the control parameter driving the system from stochastic exploration toward deterministic projection. The global motion of the dyad on the tatami is modelled as Fractional Brownian Motion (fBm) (Section 4.1.1, Eq. 7), with the local Hölder exponent h_t (Eq. 5) and the Hurst parameter H (Eq. 6–7) $H$ quantifying persistence or anti-persistence. This stochastic layer captures the tactical search process that precedes the biomechanical execution of a throw.

## 6.2 Instability Analysis

Two complementary instability archetypes—vertical rotational collapse (Uchi-mata) and gravitational collapse (Seoi-otoshi, suwari version)—are introduced as dynamical abstractions defining a continuum onto which all throwing techniques can be mapped (Section 3.3.1). These archetypes exhibit distinct dynamical signatures: Rotational archetype: progressive Lyapunov divergence, formation of rotational attractors. Gravitational archetype: discontinuous phase transitions, gravitational acceleration amplification.
The Functional Instability Index $I(t)$ (Eq. 2, Section 3.3) is an order parameter derived from local bifurcation analysis, integrating dynamical stability (μ_max), geometric proximity to the support boundary (η), and grip impedance (K_grip). It replaces the previous weighted sum of six variables. The original variables included: finite-time Lyapunov exponent $\lambda_N(t)$, normalised rotational torque $T_{\mathrm{rot},N}(t)$, asymmetry index $A(t)$, normalised COM displacement $d_N(t)$, normalised downward acceleration $a_z(t)$, normalised grip stiffness $K_{g,N}(t)$. A sigmoid soft-threshold function $S$ maps the combined instability metric χ(t) onto $[0, 1]$, where 0 indicates stable reciprocal interaction and 1 indicates irreversible collapse. (Full derivation and normalisation are provided in Appendix B.4).

## 6.3 Analytical Tools

The framework employs established tools from nonlinear time-series analysis, adapted to the transient, non-stationary nature of judo interactions:

1. Finite-Time Lyapunov Exponents (FTLE, Eq. 3): Estimated via the Rosenstein algorithm over sliding windows to capture local divergence during the Kuzushi-to-Kake transition.
2. Poincaré Maps (Eq. 4): Constructed at characteristic phases of dyadic motion (e.g., foot-plant events) to reveal attractor structures and detect interruptions in rotational or directional flow.
3. Hurst Parameter Estimation (Eq. 6): Computed from the variance of dyadic COM displacements across time scales to classify tactical regimes: stochastic exploration, persistent attack, anti-persistent defence.

4. Phase Transition Detection: Critical thresholds in Kumi-kata are identified through abrupt changes in: grip topology, relative orientation, local Hölder exponent h_t $\frac{dH}{dt}$ (Eq. 5).

### 6.4 AI Pipeline (Overview)

The translation of the theoretical framework into practical coaching tools is mediated by a multistage AI pipeline (full details in Appendix A). The pipeline proceeds through six stages:

1. Data Acquisition – High-frequency video capture (120-240 fps, single or multi-camera).
2. Pose Estimation – Deep-learning extraction of 2D/3D keypoints for both athletes, followed by segmental COM computation.
3. Dyadic Feature Extraction – Computation of symmetry indices, micro-instability markers, relative displacement, and Kumi-kata transition metrics.
4. Nonlinear Metric Computation – Estimation of FTLE, Poincaré maps, $I(t)$, local Hölder exponents h_t (Eq. 5), and Hurst parameters H (Eq. 6) from extracted trajectories.
5. Machine Learning Classification
    - Unsupervised clustering (HDBSCAN) for technique discovery.
    - Supervised calibration of the logistic threshold χ_crit and steepness k of $I(t)$using expert-labelled sequences.
6. Coaching Outputs – Generation of instability heat maps, early-warning indicators, critical-event summaries, and longitudinal athlete profiles.

The pipeline is designed to be modular: each stage can be refined independently as algorithms improve and as larger, higher-quality datasets become available. The current specification represents a blueprint for implementation, not a report of completed empirical work.

## 7. Towards a New Advanced Didactic Program for High-Level Competition

The nonlinear dynamical framework developed in this study not only reinterprets judo interactions but also provides the foundation for a structured, instability-centred teaching methodology. Traditional judo pedagogy is typically organised around techniques, biomechanical principles, or tactical scenarios. When judo is reframed as a coupled dynamical system, the primary pedagogical variable becomes instability—its perception, manipulation, and exploitation. In line with the applicative implications discussed by Lopes (2024), training should expose athletes to goal-stable regions, promote perceptual-motor exploration, and facilitate the perception of projection affordances. The three-level educational model proposed here integrates fully with this perspective.

### 7.1 The Unified Nonlinear Didactic Program: Integrating Equations 1, 2, and 7

Within the Ecological Dynamics framework (Yearby et al., 2024; Davids et al., 2015), training can be organised into three levels of complexity. The nonlinear classification developed in this study demonstrates that techniques are not isolated entities but continuous applications of a single principle: instability.
The three core equations of the model define the pedagogical structure:

1. Eq. 1 – The Tool: defines the constraint structure and the archetype to employ.
2. Eq. 7 – The Map: dictates how to navigate the stochastic field (environment).
3. Eq. 2 – The Trigger: establishes when to initiate system collapse.

Once collapse begins, Tori applies the mechanical tool (lever or couple). Artificial intelligence functions as a virtual sensor, enabling coaches and athletes to visualise these structures and transform intuitive mastery into a repeatable, quantifiable, scientifically validated training protocol.

**Level 1 – Perception of Instability**

Focus on static asymmetry detection. Athletes learn to read posture, COM deviation, and grip-induced constraints. Techniques are reclassified by their internal constraint configuration $J_q^T \lambda(t)$, transforming each throw into an optimised mechanical solution.
*Examples:* Vertical rotational progression (Uchi-mata) building a rotational attractor; Gravitational collapse (Seoi-otoshi, suwari) amplifying asymmetry to trigger instantaneous collapse.

### Level 2 – Manipulation of Instability

The second level focuses on inducing instability through dynamic symmetry breaking. Athletes learn to manipulate rhythm, timing, coupling, micro-perturbations, grip modulation, and controlled entries (Tsukuri). This generates local divergence measurable through finite-time Lyapunov exponents.
This level corresponds to the intermediate regime of $I(t)$, where dynamic contributors (FTLE, rotational torque, coupling variations) dominate.
*Stochastic Navigation and Field Perception (Eq. 7 – fBm)*
The tatami is conceptualised as a fractional Brownian field. Athletes learn to exploit the "memory" of movement (persistence) to conserve energy and use the opponent's stochastic noise to fuel projection.
Scientific Core: fBm defines the "texture" of combat. Didactic Objective: developing Hurst sensitivity.
Application: "surfing" stochastic fluctuations to position the dyad in optimal basins of attraction.

### Level 3 – Exploitation of Instability and Archetype-Specific Finalisation

At the third level, the athlete learns to exploit the instability previously created, guiding the dyad into one of the two fundamental attractors: rotational attractor (Uchi-mata archetype) or gravitational collapse attractor (Seoi-otoshi archetype). The athlete must recognise the point of no return, where the system crosses the instability threshold and enters irreversible collapse.
In the Rotational Instability Archetype: maintain flow continuity and amplify vertical-rotational torque.
In the Gravitational-Lever Instability Archetype: maximise descent velocity and remove vertical support beneath Uke's COM.
This corresponds to the upper region of $I(t)$, where instability becomes self-amplifying and the mechanical tool finalises the throw.
*Triggering the Singularity (Eq. 2 – Instability Index)*
The student learns to identify the moment of maximum instability and apply a minimal $Q_{\text{musc}}$ to trigger collapse.
Scientific Core: monitoring $I(t)$. *Didactic Objective*: singularity timing.
*Outcome*: controlled dynamic symmetry breaking, converting the hidden order of Kumi-kata into the physical reality of projection (lever or couple).

## 7.2 Summary

This nonlinear teaching framework transforms judo pedagogy from a technique-centred to an instability-centred approach. By aligning perceptual, coordinative, and tactical skills with the dynamical structure of the Tori–Uke dyad, the model provides a coherent progression reflecting the true nature of competitive interactions. AI integration strengthens this approach by offering measurable indicators that support athlete development and scientific validation. This program transforms the coach into a system analyst and the athlete into a controller of complex dynamics.

#### Practical Application: Training the Phase Transition

The most critical skill for a modern judoka is managing the transition between scales. Coaches should implement *Topological Randori*, where the goal is not the throw but achieving a specific asymmetric configuration in Uke. The AI pipeline (Appendix A) provides the objective ground truth to verify whether the athlete correctly identifies these hidden opportunities.

***Table 1 Mapping Between Nonlinear Dynamics Concepts and the Three-Level Teaching Framework***

| **Theoretical Construct** | **Description (Dynamics of the Dyad)** | **Didactic Level** | **Teaching Focus** | **Measurable Indicators** |
|---|---|---|---|---|
| **Static Symmetry Breaking** | Postural asymmetry; COM shift; locked DoF | **Level 1** | Perception of instability; reading posture | COM offset $d_N(t)$; torso angle asymmetry; grip-induced constraints |
| **Dynamic Symmetry Breaking** | Perturbation-driven divergence; rhythm-induced instability | **Level 2** | Manipulation of instability; timing and rhythm | FTLE (short window); timing mismatch; coupling variation |
| **Coupling Stiffness $K_g(t)$** | Grip tension as stabilizing/destabilizing parameter | **Level 1–2** | Understanding grip effects; modulating stiffness | Grip force; stiffness index; bilateral tension asymmetry |
| **Local Divergence (FTLE)** | Sensitivity to initial conditions; onset of chaos | **Level 2** | Entering meta-stable regions; inducing divergence | FTLE > 0 threshold; divergence rate |
| **Attractors (Rotational / Collapse)** | Stable tendencies of the dyad; phase-space structure | **Level 3** | Guiding the dyad into the correct attractor | Attractor shape; rotational continuity; descent acceleration |
| **Rotational Torque $T_{rot}(t)$** | Vertical–rotational instability (Uchi-mata archetype) | **Level 3** | Exploiting rotational instability | Torque peak; hip–shoulder alignment timing |
| **Downward Acceleration $a_z(t)$** | Gravitational collapse (Seoi-otoshi archetype) | **Level 3** | Exploiting gravitational discontinuity | Acceleration spike; drop velocity |
| **Instability Index $I(t)$** | Global measure of instability (0–1) | **Level 1–3** | Recognizing the instability window | Threshold crossing; instability growth rate |
| **Hurst Parameter (H)** | Exploration vs. deterministic attack | **Level 1** | Understanding tactical exploration | H < 0.5 (anti-persistent); H > 0.5 (deterministic) |
| **Tactical Impedance Forces Q_tact** | Tactical saturation of Uke's DoF | **Level 2–3** | Reducing Uke's phase space; controlling transitions | DoF reduction; constraint activation timing |

## 8. Broader AI Applications in Judo and Beyond

The integration of nonlinear dynamics with artificial intelligence opens a broad spectrum of applications extending far beyond the analysis of individual throwing actions. When judo is treated as a coupled dynamical system, AI becomes both a microscope for hidden instability patterns and a predictive tool for performance optimisation. The following domains represent the most promising directions for translating this theoretical framework into practical impact.

### 8.1 Tactical Pattern Recognition

The AI pipeline described in this work (Appendix A) can be extended to automated tactical scouting. By analysing large corpora of competition footage, the system identifies recurring dyadic signatures such as: preferred Hurst regimes before attacking, characteristic symmetry-breaking configurations, vulnerability windows following specific Kumi-kata transitions.

Coaches can query the system with questions such as: *"When does Athlete (X) most frequently enter the rotational instability archetype?"* The system provides data-driven answers grounded in nonlinear metrics rather than subjective observation. This transforms match preparation from qualitative video review into quantitative analysis of dynamical tendencies.

**8.2 Match Analysis Based on Nonlinear Dynamics**

Traditional match analysis quantifies discrete events (attacks, grip breaks, scores). The proposed framework enables a fundamentally different approach: continuous monitoring of the dyad's stability landscape throughout the match.

By computing the Functional Instability Index $I(t)$ over full competition time series, analysts can generate instability profiles that reveal: proximity to phase transitions, meta-stable configurations, persistent dominance in Hurst space, near-threshold fluctuations invisible to conventional statistics. A match that appears uneventful in terms of scores may exhibit rich dynamical structure that predicts future outcomes more reliably than traditional metrics.

### 8.3 Injury-Risk Modelling

Nonlinear dynamics offers a novel lens for injury prevention. High-injury scenarios often involve unexpected collapses, counter-throws, or landings where instability exceeds the athletes' capacity for controlled decoupling. AI can identify dangerous transitions by detecting: Lyapunov exponents exceeding safe thresholds during Sutemi-waza, asymmetry indices indicating loss of controlled landing posture, instability spikes preceding counter-throws.

Over longitudinal tracking, an athlete's instability footprint may reveal fatigue-related degradation of control, providing an early-warning system for overuse or acute injury risk before clinical symptoms appear.

### 8.4 Automated Technical Classification

The instability-based classification proposed in Section 3.4—mapping all throwing techniques onto a continuum between rotational and gravitational collapse archetypes—naturally lends itself to automated labelling.

Unsupervised clustering algorithms (e.g., HDBSCAN; Appendix A.5) can: organise large datasets of competition throws by dynamical signature, reveal hybrid techniques not captured by traditional nomenclature, validate or refine existing classification systems, provide objective criteria for distinguishing new techniques from stylistic variants. This shifts classification from static labels to dynamic signatures.

### 8.5 Cross-Domain Extensions

Although developed for judo, the core architecture—coupled dyadic systems, symmetry breaking, instability indices, stochastic navigation via fBm—is transferable to other grappling and combat sports, including: Brazilian Jiu-Jitsu, freestyle and Greco-Roman wrestling, Sambo, striking sports with coupled timing (boxing, fencing).

Beyond sport, the mathematical structure may inform: human-robot physical interaction, rehabilitation robotics, cooperative manipulation tasks, any domain involving two agents interacting through physical coupling and tactical intention.

### 8.6 AI-Based Match Analysis Pipeline (Summary)

The complete pipeline for operationalising these applications follows the architecture detailed in Appendix A and proceeds through six stages:

1. High-frequency video acquisition (120-240 fps).
2. Pose estimation.
3. Dyadic feature extraction (symmetry indices, micro-instability markers, grip topology transitions).
4. Nonlinear metric computation (FTLE, Poincaré maps, Instability Index).
5. Machine learning clustering and classification.
6. Coaching outputs (instability heat maps, early-warning indicators, longitudinal profiles).

AI applied to judo must operate on properly detrended signals. As highlighted by Bryce & Sprague (2012), DFA introduces artefacts in the presence of nonlinear trends, compromising feature quality. This confirms the need for AI pipelines based on physically interpretable measures rather than biased scalability estimates. This pipeline bridges the gap between the conceptual framework and actionable, data-driven coaching tools.

**Table 2 – Stages of the Proposed Pipeline**

| Step | Process | Technical content |
|---|---|---|
| 1 | **Input Video** | Video capture of the match; single or multi-camera. |
| 2 | **Pose Estimation** | Detection of the two athletes' keypoints; reconstruction of the dyad configuration. |
| 3 | **Dyadic Features Extraction** | - Symmetry indices / symmetry breaking - Micro instabilities - Relative displacement - Kumi kata transitions |
| 4 | **Non-linear Metrics** | - Trajectory Divergence - Critical Thresholds - Nonlinear Amplification - Instability Classification |
| 5 | **Tactical Interpretation Layer** | - Windows of vulnerability - Tactical opportunities - Recurring adversary patterns |
| 6 | **Output: Dynamic Match Report** | Summary of critical events, detected instabilities, tactical opportunities and overall dynamics of the encounter. |

## 9. Conclusion

This foundational theoretical work applies nonlinear dynamics to competitive judo, enabled by recent advances in artificial intelligence that allow the evaluation of transient situations imperceptible to the human eye. Viewing competition as a dynamic continuum shifts the classical perspective from a catalogue of throwing techniques to the more subtle and decisive task of generating instability within the dyad. In this view, competition becomes the systematic creation of static or dynamic instability in the opponent to enable the application of the two fundamental mechanical tools of projection: Couple or Lever. Paradoxically, in Ne-waza (ground fighting), the logic reverses: while standing techniques aim to create instability, osaekomi requires the progressive construction of maximum stability. This duality reinforces the central role of stability–instability dynamics across all phases of judo.
This framework demonstrates that judo excellence lies in the mastery of instability. By integrating nonlinear dynamics with AI, coaching moves beyond qualitative intuition toward a predictive science of fight. Recent research (Dindorf et al., 2024; Rojas-Valverde, 2025) confirms that real-time quantification of Lyapunov exponents and entropy shifts will soon allow coaches to evaluate the mechanical quality of an attack before its completion. This work therefore sets the stage for a digital transformation of judo, where the intuition of the Master is supported by the mathematical structure of complex systems (Appendix B).
We propose a unified conceptual and methodological framework combining nonlinear dynamics and AI to expose the hidden structure of instability in judo competitions. By reframing techniques as dynamical transitions between basins of attraction across critical thresholds and via symmetry breaking, we identified two complementary archetypes:

1. Vertical Rotational Collapse (e.g., Uchi-mata),
2. Gravitational Collapse (e.g., Seoi-otoshi, suwari version).

These archetypes capture the fundamental pathways to projection. All throwing techniques can be viewed as points on a manifold stretched between these two poles of instability. Sutemi-waza techniques cluster toward the Seoi-otoshi pole (maximal gravitational instability), while Ashi-waza techniques involving pure rotation gravitate toward the Uchi-mata pole. This instability-based classification is particularly significant because it remains independent of athlete morphology, focusing exclusively on the physics of the competitive event.
Crucially, transitions along this manifold are governed by the Hurst parameter $H$, which quantifies the informational persistence of Kumi-kata. As $H$ shifts from stochastic noise ($H \approx 0.5$) to a persistent signal ($H \to 1$), the system is driven across a critical threshold $\theta$. At this singularity, the informational phase ends and the mechanical execution begins through the application of the two fundamental physical tools: Levers for linear amplification and Couples for angular momentum transfer.
Quantitative dynamical measures—Lyapunov exponents, Poincaré analysis, basin mapping—and the AI-based video pipeline (Appendix A) make these concepts operational. They enable: objective detection of micro-timing windows, classification of archetypal signatures, longitudinal tracking of athlete-specific attractors such as Tokui-waza.
While this manuscript serves as a foundational theoretical contribution, robust empirical validation using competition datasets remains essential. Future studies should apply the pipeline to diverse datasets, perform sensitivity analyses across recording conditions and weight classes, and evaluate the practical impact of AI-guided training through controlled trials. Additional research directions include: injury-risk identification, referee support and scoring objectivity, transfer to other grappling sports, integration into long-term athlete development programs. Careful attention to data quality, athlete consent, and reproducibility will be critical for translating these theoretical insights into reliable coaching tools and evidence-based practice.
For a step-by-step illustration of the Instability Index applied to a real competition sequence, see Appendix C.

## References (APA 7)

## Glossary of Nonlinear Dynamics for Judo Coaches

A Practical Guide to Essential Dynamical Concepts for Understanding Instability in Competition

**Attractor**

A configuration or state toward which the system "naturally tends." In judo, every technique has its attractor: the position of maximum imbalance from which Uke can no longer recover.

On the tatami: Tori's task is not to "apply a technique" but to guide the dyad toward the specific attractor of that throw. When you reach it, the fall becomes inevitable.

Example: In Uchi-mata, the attractor is a rotational configuration in which Uke's body is tilted forward and rotating, at first around his centre of mass beyond the base of support

**Basin of Attraction**

The set of all starting positions from which the system will inevitably end up in a given attractor. The wider the basin, the more "robust" the technique (it works even if execution is not perfect).

On the tatami: Techniques with a wide basin (e.g., a well-set O-soto-gari) forgive small errors. Techniques with a narrow basin (e.g., dropping Seoi-otoshi) demand millimetric timing.

Coaching implication: Teach techniques with wide basins first to build confidence; then refine toward narrower basins for high-level competition.

**Bifurcation**

The critical moment when the system can take two completely different paths. In judo, it is that point where an extremely small variation decides whether the attack succeeds or fails.

On the tatami:The bifurcation is the "point of no return." Before it, Uke can still defend. After it, fate is sealed.

Practical signal: When you see Uke stop making micro-corrections and "accept" the direction imposed by Tori, you have entered the bifurcation zone.

**Coupling / Coupled System**

Two athletes in grip are not two separate bodies: they form a single system in which each movement of one instantly modifies the other's possibilities. This is what we call "the Dyad."

On the tatami: You cannot think only about your own action. You must "feel" how Uke reacts and adapts in real time. The grip (Kumi-kata) is the channel through which this connection lives.

Coaching implication: Training grip sensitivity is as important as training strength. An athlete who "feels" the coupling can anticipate the opponent's intentions.

**Degrees of Freedom – DoF**

The number of independent directions or movements an athlete can perform. The more degrees of freedom Uke has, the more defensive options they possess.

On the tatami: Tori's tactical objective is to saturate Uke's degrees of freedom: close spaces, limit escape directions, constrain movements through grips. When Uke has zero functional degrees of freedom, they are ready to be thrown.

Practical rule: If Uke can still freely move their head, hips, or pivot foot, they still have available degrees of freedom. The attack is premature.

**Deterministic / Stochastic**

A deterministic process is predictable: given the cause, the effect necessarily follows. A stochastic process contains randomness or noise. The Kumi-kata phase is predominantly stochastic (exploration, continuous adaptation); the Kake phase is predominantly deterministic (the fall follows physical laws).

On the tatami: Do not attack during the stochastic phase. Wait until you have reduced the "noise" and created a condition in which your action produces a predictable effect.

Key transition: The passage from stochastic to deterministic is the heart of Tsukuri. It is the moment you stop "searching" and start "executing."

**Divergence**

Two nearly identical situations that evolve in completely different ways. In judo, this explains why the very same movement can succeed in one instant and fail a moment later.

On the tatami: Divergence is amplified in unstable systems. Your task is to create conditions in which divergence plays in your favour: a small push produces a large imbalance.

Implication for video analysis: When reviewing a failed attack, do not look only for the gross error. Look for the micro-difference (an angle of 3°, a delay of 50 ms) that made the system diverge toward failure.

**Entropy**
A measure of disorder or uncertainty in the system. High entropy = many possibilities, low predictability. Low entropy = few possibilities, high predictability.
On the tatami: At the start of a bout, entropy is high: both athletes have many options. A good Kumi-kata strategy progressively reduces Uke's entropy, "freezing" their options until only one path remains: falling.
Tactical objective: Your action must reduce the opponent's entropy faster than they reduce yours.
**Finite-Time Lyapunov Exponent – FTLE**
A number that measures how quickly two nearly identical trajectories separate over time. If positive, the system is unstable and small causes produce large effects. If negative or zero, the system is stable.
On the tatami: This is the "instability growth rate." When FTLE becomes positive and grows rapidly, you are in the Kake window: every micro-action by Tori is amplified into a large imbalance in Uke.
For video analysis: With high-speed cameras (240 fps), AI can compute FTLE and tell you exactly in which frame the instability began to grow irreversibly.
**Fractional Brownian Motion – fBm**
A mathematical model describing how the couple of athletes moves on the tatami. Unlike a random walk, fBm has "memory": past movements influence future ones.
On the tatami: Movement on the tatami is not random. If Tori is dominating tactically, the movement will show persistence (continuing in the same direction). If Uke is reacting well, it will show anti-persistence (frequent direction changes).
Tactical reading: An opponent moving with a persistent pattern is likely "committed" in one direction and easier to surprise with a change.
**Hurst Parameter – H**
A value between 0 and 1 that quantifies the "memory" of movement. $H = 0.5$: random movement (no memory). $H > 0.5$: persistence (movement continues in the current direction). $H < 0.5$: anti-persistence (movement tends to reverse direction frequently).
On the tatami: Monitoring H during a bout tells you who is controlling the rhythm. An H rising toward 1 indicates Tori is imposing their direction. An H falling toward 0 indicates Uke is making continuous defensive micro-corrections.
Coaching implication: Train your athletes to recognise the "texture" of the opponent's movement. An opponent with $H < 0.5$ is in reactive mode and can be led into making mistakes.
**Instability Index – I(t)**
A number from 0 to 1 measuring how close the Tori-Uke system is to collapse. 0 = perfectly stable. 1 = inevitable fall. It integrates geometry, dynamics, forces, and grip into a single measure.
On the tatami: The objective is not to apply brute force but to raise I(t) as efficiently as possible. An I(t) value around 0.7–0.8 is the "golden window" for initiating Kake. Earlier is too soon; later is too late.
Future application:
With AI, I(t) could be estimated in real time from video, providing objective feedback on attack quality before the throw is even completed.
**Multibody System**
A mathematical model describing a system composed of multiple rigid bodies connected by constraints. The Tori-Uke dyad in grip is a multibody system with physical constraints (foot-mat contact, grips) and tactical constraints (saturation of degrees of freedom).
On the tatami: When you grip Uke, you are not just pulling a sleeve: you are creating a complex mechanical structure in which your levers and constraints determine the movement possibilities of both athletes.
Coaching implications: Teaching grips not as "grabbing" but as "building a constrained structure" will change how athletes understand Kumi-kata.
**Nonlinear Amplification**
In a linear system, a small input produces a small output. In a nonlinear system, a small input can produce a huge and disproportionate output. Judo is nonlinear: a small symmetry break can generate an Ippon.
On the tatami: You do not need to be stronger. You need to apply the right force, at the right moment, in the right direction, when the system is in the configuration that amplifies that input.
Key teaching: Perfect technique is not the one with maximum force, but the one with maximum amplification.

**Phase Space**
An abstract space in which each point represents the complete state of the Dyad system (positions, velocities, orientations of both athletes). The trajectory in this space tells the entire story of the interaction.
On the tatami: You cannot see phase space with the naked eye, but you can "feel" it. When an opponent is in a position from which they have few options, they are in a "narrow" region of phase space. That is where you must attack.
With AI: Artificial intelligence can reconstruct and visualise phase space, showing coaches the regions where the athlete is most vulnerable or most effective.
**Phase Transition**
A sudden and qualitative change in the state of the system. Just as water turns to ice, the dyad passes from "stable" to "falling" through a phase transition.
On the tatami: The phase transition is the passage from Tsukuri to Kake. It is not gradual: it is a jump. Recognising the exact moment of transition is what distinguishes a champion from a good athlete.
Practical indicator: When you feel Uke's resistance suddenly "give way" and the system flows without further obstacles, you have crossed the phase transition.
**Poincaré Map**
A "periodic snapshot" of the system that reveals the hidden structure of movement, like observing a dancer always at the same point in the choreography. It reduces complexity by revealing repetitive patterns.
On the tatami: Taking "snapshots" of the system at key moments (e.g., every foot plant) allows you to see whether the movement is evolving in the desired direction or whether there are interruptions in the flow of the attack.
For video analysis: AI can automatically generate Poincaré maps to compare different executions of the same technique and identify where fluidity breaks down.
**Saddle Point**
A point of unstable equilibrium: the system can stay there, but the slightest perturbation will make it fall one way or the other. It is the "ridge" between success and failure.
On the tatami: The saddle point is the moment of maximum uncertainty in the attack: Uke is teetering, and a small push will decide whether they fall or recover. Great champions recognise and exploit this instant.
Coaching implication: Train your athletes to "feel" the saddle point and to apply the decisive impulse exactly at that instant.
**Self-Organisation**
The system's capacity to spontaneously form coordinated patterns without central control. In judo, the dyad self-organises: the movements of Tori and Uke coordinate into patterns that neither planned individually.
On the tatami: You cannot control everything. You must learn to "let the system do its work" once you have set the right initial conditions.
Coaching implication: Free Randori is essential precisely because it trains this capacity for adaptation to the system's self-organisation.
**Stability Landscape**
A visual metaphor: imagine a surface with valleys (stable zones) and hills (unstable zones). The dyad moves across this landscape. Attacking means pushing the system out of a valley and toward a descent leading to **Ippon.**
On the tatami: Your opponent is in a "valley of stability." Kuzushi consists of bringing them to the ridge. Kake is the descent on the other side.
For coaches: Visualising the stability landscape helps explain why certain positions are more dangerous than others, even if they appear similar.
**Symmetry / Symmetry Breaking (Sacripanti)**
Symmetry is the perfect balance between Tori and Uke (both stable, same degrees of freedom). Symmetry breaking is the moment this balance is disrupted in favour of one of the two.
On the tatami: Two types exist: (1) Static – when Uke already has an asymmetrical posture (e.g., rotated torso, weight on one leg); (2) Dynamic – when asymmetry is created by Tori's movement and timing. Recognising static symmetry gives you a target; creating dynamic symmetry is the essence of Kuzushi.
Practical rule: If Uke is symmetrical and balanced, do not attack with force: first break the symmetry.
**Tactical Impedance Control (formerly: Virtual Constraints)**
Unlike physical constraints (grip, the ground), virtual constraints are "tactical impositions" that limit the opponent's options: space management, rhythm, direction of movement, attack threat.

On the tatami: When you "close the space" or "cut the escape direction" with body positioning, you are modulating K_grip(t) and D_grip(t) — the stiffness and damping of the tactical coupling. There is no physical wall, but Uke cannot go there without exposing themselves.
Coaching implication: Teach footwork not as "moving" but as "building tactical impedance around Uke."

***Note for coaches***: This glossary is a translation tool. You do not need to memorise the mathematical definitions, but to assimilate the concepts through experience on the tatami. Every time you observe a sudden transition, a small action producing a large effect, or a moment when "everything flows," you are observing nonlinear dynamics in action. AI is your new eye for seeing what experience alone could only intuit.

# APPENDIX A

### Detailed AI Pipeline for Operationalising Nonlinear Dynamics in Judo

The transition from qualitative coaching observation to quantitative dynamical analysis is mediated by a multistage AI pipeline designed to process high-dimensional, noisy data from elite judo competitions. This appendix provides the full technical specification of the pipeline referenced throughout the manuscript.

## A.1 Data Acquisition and Pre-processing

High-frequency capture: Raw video is ideally recorded at 120-240 fps to resolve "micro-times" during the transition from Tsukuri to Kake. Geometric calibration: For multi-view setups, Direct Linear Transformation (DLT) or equivalent algorithms map 2D coordinates into a unified 3D global space centred on the tatami.

## A.2 3D Pose Estimation and Biomechanical Reconstruction

Skeletal extraction: Deep-learning architectures (e.g., Stacked Hourglass Networks, HRNet) extract 2D/3D coordinates for 17-25 keypoints per athlete. COM calculation: Global centres of mass (COM) for Tori and Uke are estimated using weighted segmental mass models. Contact-tension proxies: Micro-variations in distance and relative acceleration between grip points and Uke's torso serve as mathematical proxies for grip stiffness or coupling strength.

## A.3 Feature Extraction for Symmetry Analysis

Asymmetry indicators: The system computes deviations of Uke's COM from the central axis of the base of support (BOS), quantifying Static Symmetry Breaking (Section 2). η(q) — norm. COM distance to support boundary vs. vertical drop: AI distinguishes between accumulation of angular momentum (rotational archetype) and rapid increases in downward acceleration (gravitational archetype).

## A.4 Dynamical Modelling (Core Engine)

Lyapunov exponent estimation: The Rosenstein algorithm is applied to reconstructed COM phase-space trajectories. A shift from negative to positive values ($\lambda > 0$) automatically signals the onset of Kuzushi. Poincaré mapping: Stroboscopic "slices" of motion at specific phases of the fBm displacement reveal underlying attractors corresponding to athlete-specific Tokui-waza.

## A.5 Machine Learning Clustering and Classification

Unsupervised discovery: Hierarchical Density-Based Spatial Clustering (HDBSCAN) identifies clusters of techniques without prior labels, enabling discovery of hybrid archetypes not captured by traditional nomenclature. Expert calibration: Clusters are mapped back to the theoretical archetypes (rotational vs. gravitational) to generate actionable feedback.

## A.6 Coaching Outputs and Visualisation

Instability maps: Temporal heat maps showing when and where stability is lost. Early-warning system: By monitoring "flickering" in the Lyapunov exponent, the AI can predict a successful throw 200-300 ms before Ippon—an actionable defensive window.

## Implementation and Estimation from Video (Practical Steps)

1. Estimate $I(t)$ from pose data: compute COM, joint angles, FTLE over sliding windows $\Delta t$; rotational torque proxies from relative angular velocities and segment inertias; grip stiffness proxies from hand-displacement variance.
2. Estimate $Q_{\text{musc}}(t)$ indirectly via acceleration magnitudes and segment inertias (inverse dynamics with estimated contact forces), or via EMG when available.
3. Estimate $\lambda_{\text{tact}}(t)$ from grip geometry: relative hand positions, grip contact-area proxies, short-term stiffness from local compliance to micro-perturbations.
4. Estimate $S(t)$ as the log-determinant of the short-window covariance matrix Σ of dyadic state variables (COMx, COMy, orientation, key joint angles): $S \propto \frac{1}{2}\log\,(\det\,\Sigma)$.
5. Calibrate coefficients $(\alpha, \beta, \gamma, \delta, \epsilon, \zeta)$ in Eq. 2 via supervised learning on labelled windows (Kuzushi/Tsukuri/Kake), using cross-validation and regularisation. Confidence intervals are obtained via bootstrap.

# APPENDIX B

# Multibody Dynamics of the Tori–Uke Dyad and Formal Definition of the Functional Instability Index

## B.1. Lagrangian Formulation of the Dyad

Let $q \in \mathbb{R}^n$be the vector of generalized coordinates describing the Tori–Uke dyad (segment angles, pelvis translation, trunk orientation, grip coordinates, etc.).

The Lagrangian is defined as:

$$L(q, \dot{q}) = T(q, \dot{q}) - V(q)$$

where: **Kinetic energy**

$$T(q, \dot{q}) = \frac{1}{2} \dot{q}^T M(q) \dot{q}$$

with $M(q)$the configuration-dependent mass matrix of the coupled system.

**Potential energy**

$$V(q) = \sum_{i=1}^{N} m_i \, g \, z_i(q)$$

Partial sensitivity with respect to each component: $z_i(q)$is the vertical position of segment $i$.
This formulation treats the dyad as a **single multibody system**, consistent with the idea that Tori and Uke form a coupled entity during Tsukuri–Kake.

## B.2. Unconstrained Euler–Lagrange Equation

The Euler–Lagrange equation is:

$$\frac{d}{dt}\left(\frac{\partial L}{\partial \dot{q}}\right) - \frac{\partial L}{\partial q} = Q_{\text{musc}}$$

Computing the derivatives yields the standard multibody dynamics form:

$$M(q)\ddot{q} + C(q, \dot{q})\dot{q} + G(q) = Q_{\text{musc}}$$

where: $C(q, \dot{q})\dot{q}$collects Coriolis and centrifugal effects (Christoffel-based formulation),

$G(q) = \frac{\partial V}{\partial q}$is the gravity vector,

$Q_{\text{musc}}$are generalized muscular torques.

## B.3. Physical and Tactical Constraints

During competition, the dyad is subject to two classes of constraints.

### B.3.1 Physical constraints

These include: foot–tatami unilateral contact, bilateral grip constraints, joint limits, segmental collision avoidance. Tori's grip action is modelled as a variable-impedance coupling. The position error in operational space (Ma-ai encoding) is:

$$h_{\text{phys}}(q) = 0$$

The generalised tactical force vector (operational space impedance control) is then:

$$J_{\text{grip}}(q) = \frac{\partial x_{\text{grip}}}{\partial q}$$

### B.3.2 Tactical (virtual) constraints

These represent **intentional actions** by Tori that reshape Uke's available phase space: modulation of grip geometry, closure of lateral or frontal space, saturation of Uke's degrees of freedom, (Makikomi, Tai Atari), directional steering of the dyad, thechniques rotational application.
They are expressed as:

$$e_{\text{x}}(q,t) = x_{\text{Uke}}(q) - x_{\text{Tori}}(q)$$

$$Q_{\text{tact}}(q,\ddot{\text{t}},t) = J_{\text{grip}}^{T}\left[F_{\text{act}}(t) - K_{\text{grip}}(t)e_{\text{x}}(q,t) - D_{\text{grip}}(t)\dot{e_{\text{x}}}(q,\ddot{\text{t}},t)\right]$$

with Jacobian:

$$J_{\text{grip}}(q) = \frac{\partial x_{\text{grip}}}{\partial q}$$

This formulation does **not** introduce rigid constraint singularities. Instead it *captures the variable-impedance dynamic coupling* of Kumi-kata: a physically bounded, modulated interaction with time-varying stiffness $K_{grip}(t)$ and damping $D_{grip}(t)$.

### B.3.3 Constrained multibody equation

The kinematic Jacobian at grip contact points is defined as:

$$J_{\text{grip}}(q) = \frac{\partial x_{\text{grip}}}{\partial q}$$

The Jacobian eigenvalue stability index (maximum real part) is:

$$\mu_{\max}(t) = \max_{i}\text{Re}\left(\lambda_{\text{i}}\big(A(t)\big)\right)$$

The constrained dynamics become:

$$M(q)\ddot{q} + C(q,\dot{q})\dot{q} + G(q) = Q_{\text{musc}} + Q_{\text{tact}}(q,\ddot{\text{t}},t) + J_{\text{phys}}^{T}\lambda_{\text{phys}}$$

This formalizes the dyad as a **constrained multibody system** in which tactical actions are modelled as *variable-impedance generalised forces* $Q_{tact}$ rather than rigid Lagrange multipliers (virtual constraints), eliminating the associated mathematical singularity and more faithfully representing Kumi-kata as a dynamic coupling with time-varying stiffness and damping.

#### B.4. Analytical Derivation of the Revised Instability Index I(t)

The previous heuristic formulation of I(t) as a weighted linear sum of six heterogeneous variables (FTLE, rotational torque, asymmetry, COM offset, downward acceleration, grip stiffness) is here replaced by an analytically derived order parameter obtained from local bifurcation analysis of the multibody system .

### B.4.1 Local Stability: Jacobian Eigenvalue Analysis

Linearise the system (Eq. 1) around the instantaneous state $x_0 = (q_0, \dot{q}_0)$. The perturbed dynamics $\delta x = (\delta q, \delta\dot{q})^T$ satisfy:

$\delta\dot{x}(t) = A(q, \dot{q}, t)\, \delta x(t)$

where $A(t) \in \mathbb{R}^{2n \times 2n}$ is the system Jacobian (see B.3.3). The maximum real eigenvalue:

$$\mu_{\max}(t) = \max_i \mathrm{Re}\left(\lambda_i\big(A(t)\big)\right)$$

If $\mu_{\max}(t) > 0$, the system exhibits local exponential divergence (related to FTLE). If $\mu_{\max}(t) < 0$, trajectories return to the quasi-equilibrium.

### B.4.2 Geometric Stability: Normalised COM Distance

The normalised distance of the projected COM from the support polygon boundary ∂B:

$$\eta(q) = 1 - \frac{\left\| x_{\mathrm{COM}}(q) - x_{\mathrm{BOS,center}} \right\|}{R_{\mathrm{BOS}}(\theta)}$$

where $R_{BOS}(\theta)$ is the radius of the support polygon in direction θ. $\eta(q) \to 1$ indicates full stability; $\eta(q) \to 0$ indicates imminent geometric collapse.

### B.4.3 Combined Instability Metric χ(t)

The dynamical-geometric combined instability metric integrates $\mu_{max}$, η, and grip impedance $K_{grip}$ multiplicatively:

$$\chi(t) = \frac{\mu_{\max}(t)}{\eta\big(q(t)\big)} \cdot \frac{K_0}{K_{\mathrm{grip}}(t) + K_0}$$

Interpretation: instability increases with eigenvalue divergence rate ($\mu_{max}$ ↑), diverges hyperbolically as the COM approaches the support boundary ($\eta \to 0$), and is mitigated by Uke's defensive grip stiffening ($K_{grip}$ ↑ suppresses χ via the factor $K_0/(K_{grip} + K_0) \to 0$).

### B.4.4 Revised Instability Index I(t) — Logistic Mapping

The logistic mapping onto [0,1] yields the order parameter:

$$I(t) = \frac{1}{1 + e^{-k(\chi(t) - \chi_{\mathrm{crit}})}}$$

where $k > 0$ is the steepness (calibrated by supervised learning on labelled competition sequences) and $\chi_{crit}$ is the critical threshold (bifurcation point). $I(t) = 0$: stable reciprocal interaction. $I(t) = 1$: irreversible collapse (Kuzushi/Kake window). $I(t) \approx 0.5$ corresponds to $\chi(t) = \chi_{crit}$, the metastable transition zone.

I(t) is thus a true order parameter derived from local bifurcation analysis, integrating dynamical stability ($\mu_{max}$), geometric stability (η), and grip impedance coupling ($K_{grip}$) in a multiplicatively nonlinear structure — replacing the previous additive heuristic.

### B.4.5 Stochastic Coupling Function $\Gamma(h_t)$

The energy flow equation (Eq. 5) introduces the coupling function that links the local Hölder exponent $h_t$ to the direction of stochastic energy injection:

$$\Gamma(h_{\mathrm{t}}) = \tanh\big(\kappa(h_{\mathrm{t}} - 0.5)\big)$$

If $h_t > 0.5$: system is persistent ($\Gamma > 0$, Tori accumulates directional dominance). If $h_t < 0.5$: system is anti-persistent ($\Gamma < 0$, Uke absorbs energy via rapid micro-corrections). The transition $\Gamma(h_t) = 0$ at $h_t = 0.5$ marks the boundary between stochastic regimes, acting as a precursor to the bifurcation captured by $\chi_{crit}$.

## B.5. Sensitivity Analysis

The Functional Instability Index $I(t)$integrates geometric, dynamic, and coupling-combines dynamical stability ($\mu_{max}$), geometric stability (η), and grip impedance ($K_{grip}$) through a multiplicatively nonlinear structure. A formal sensitivity analysis is required to understand how variations in each component affect the output, particularly near the critical bifurcation threshold $\chi_{crit}$. The formulation is:

I(t) = 1/(1+e^{−k(χ(t)−$\chi_{crit}$)}), where χ(t) = $\mu_{max}$(t)/η(q(t)) · $K_0$/($K_{grip}$(t)+$K_0$). Because χ is a ratio of dynamical and geometric quantities modulated by grip impedance. A formal sensitivity component influence the global instability estimate.

### B.5.1 Mathematical Structure of the Revised Index

Let:

$$\chi(t) = \frac{\mu_{\max}(t)}{\eta(q(t))} \cdot \frac{K_0}{K_{\text{grip}}(t)+K_0}$$

Then:

$$I(t) = \frac{1}{1 + e^{-k(\chi(t) - \chi_{\text{crit}})}}$$

### B.5.2 Local Sensitivity of χ(t) (Partial Derivatives)

$$\frac{\partial I}{\partial \mu_{\max}} = \frac{S'(\chi)}{\eta(q)} \cdot \frac{K_0}{K_{\text{grip}}(t) + K_0}$$

where $\frac{\partial I}{\partial \eta} = -\frac{S\prime(\chi)\cdot\mu_{\max}(t)}{\eta^2} \cdot \frac{K_0}{K_{\text{grip}}(t)+K_0}$The sensitivity of I to each component of χ: $S'(\chi) = k \cdot I(t) \cdot \big(1 - I(t)\big)$The logistic activation derivative: S′(χ) = k · I(t) · [1−I(t)], which is maximal at χ = $\chi_{crit}$ (metastable Kuzushi window).

$$S'(\chi) \;=\; k \cdot I(t) \cdot [1 - I(t)]$$

**Biomechanical interpretation of each component:**

### B.5.3 Normalised Sensitivity Index for χ(t)

$$S_{\text{i}}(t) = \left|\frac{\partial I}{\partial x_{\text{i}}} \cdot \frac{x_{\text{i}}}{I(t)}\right|$$

The normalised sensitivity index S_i = |∂I/∂x_i · x_i/I(t)| measures relative influence. Expected patterns:
**Uchi-mata:** high sensitivity to $\mu_{max}$ (Jacobian eigenvalue divergence drives χ via the numerator);

**eoi-otoshi:** high sensitivity to η (proximity to support boundary drives χ hyperbolically as η→0);

**Hybrid techniques:** shifting sensitivities

### B.5.4 Global Sensitivity (Parameter Variation)

$$\partial I/\partial Kgrip \;=\; -S'(\chi) \cdot \mu_max/\eta \cdot K_0/(Kgrip + K_0)^2$$

Large $K_{grip}$ relative to $K_0$ (e.g., strong defensive grip in $\chi_{\text{crit}} \to 0$in Seoi-otoshi) strongly dampens χ and hence I(t). Conversely, $K_{grip} \to 0$ (grip loss) makes the impedance factor approach 1, maximising sensitivity to μ_max/η.

### B.5.5 Zones of Maximum and Minimum Sensitivity of I(t)

- **Maximum:**
  - $I(t) \in [0.4, 0.7]$
  - $\mu_{max} > 0$, $\eta \to 0$ (COM near support boundary), $K_{grip} \to 0$ (grip loss)

- **Minimum:**
  - $I(t) < 0.2\mu_{max} < 0$ (stable basin)
  - $I(t) > 0.85$I(t) > 0.85 (irreversible collapse)
  - $\chi_{\text{crit}} \to 0K_{grip} >> K_0$ (strong defensive coupling, χ suppressed)

### B.5.6 Summary Table (Revised Index Components)

**Summary Table**

| Component | Meaning | Sensitivity behaviour | Dominant Archetype |
|---|---|---|---|
| $A_N$ | μ_max(t) — max real Jacobian eigenvalue | High: drives χ numerator; diverges near bifurcation- | Uchi-mata (rotational)- |
| $T_{\text{rot}}$ | η(q) — norm. COM distance to support boundary | Very high near collapse (η→0, χ→∞) | Seoi-otoshi (gravitational)- |
| $A$ | K_grip(t) — grip impedance stiffness | Stabilising: K_grip↑ suppresses χ; K_grip→0 amplifies it | Both archetypes |
| $d_N$ | χ(t) = μ_max/η · K₀/(K_grip+K₀) | Multiplicative: all three components interact nonlinearly | Both archetypes- |
| $a_z$ | | | - |
| $K_g$ | | | |

### B.5.7 Conclusions

The sensitivity analysis of the revised I(t) demonstrates that: the index is most responsive in the **Tsukuri–Kake transition**; each archetype exhibits a distinct sensitivity signature in μ_max vs η dominance; the logistic activation on χ ensures numerical stability at saturation; the impedance term K₀/(K_grip+K₀) provides smooth, bounded sensitivity to grip coupling; the new multiplicative structure avoids the redundancy between FTLE and kinematic variables present in the old linear sum. This analysis strengthens the theoretical foundation of the Instability Index as a meaningful order parameter for the dyadic system.

## B.6 Mathematical Characterisation of the Two Instability Archetypes

### B.6.1 Notation

Let:

a. $C_U(t)$: Uke's centre of mass

b. $C_T(t)$: Tori's centre of mass
c. $B_U$: Uke's support polygon
d. $\vec{F}(t)$: resultant force applied by Tori
e. $\vec{\tau}(t)$: torque acting on Uke
f. $F(t)$: time-dependent fulcrum (feet → shins)
g. $\mathbf{I}_{F(t)}$: inertia tensor of Uke with respect to the instantaneous fulcrum
h. $\mathbf{R}(t)$: rotation matrix around the instantaneous axis
i. $\vec{d}(t)$: downward translation induced by gravity and Tori's drop

Irreversible instability occurs when:

$$I(t) > 1, \frac{dI}{dt} > 0.$$

### B.6.2 Rotational Instability Archetype (Uchi-mata Type)

Instability is dominated by a torque around an oblique axis $\hat{k}_R$. Let $\vec{r}(t)$be the lever arm from the COM to the point of force application.

$$\vec{\tau}_R(t) = \vec{r}(t) \times \vec{F}(t),$$
$$\mathbf{I}_U\,\dot{\vec{\omega}}(t) = \vec{\tau}_R(t),$$
$$\hat{k}_R(t) = \frac{\vec{\tau}_R(t)}{\parallel \vec{\tau}_R(t) \parallel}.$$

The COM trajectory is:

$$C_U(t) = C_U(0) + \int_0^t \mathbf{R}(\tau)\ \vec{v}(\tau)\,d\tau.$$

Instability becomes irreversible when:

$$\text{proj}_{\,B_U}(C_U(t)) \notin B_U.$$

### B.6.3 Mixed Gravitational–Lever Instability Archetype (Seoi-otoshi, Suwari Version)

This archetype combines:

1. a gravitational collapse (rapid drop of Tori's COM), and
2. a lever action with a migrating fulcrum.

The shoulder is a **kinematic constraint**, not a fixed pivot.

### B.6.3.1 Time-dependent fulcrum

$$F(t) = \{F_1, \quad 0 \le t < t_c\text{(feet–tatami friction dominates)}\ \ F_2, \quad t \ge t_c\text{(shins on Tori friction dominate)}$$

Torque about the moving fulcrum:

$$\vec{\tau}_F(t) = \vec{r}_{F(t)\to C_U}(t) \times \vec{F}(t),$$
$$\mathbf{I}_{F(t)}\,\dot{\vec{\omega}}_F(t) = \vec{\tau}_F(t).$$

### B.6.3.2 Helical descent of the COM

Because the fulcrum moves and the applied force has both vertical and lateral components, the COM follows a **descending helical trajectory**:

$$C_U(t) = F(t) + \mathbf{R}(t)\,\vec{\rho} + \vec{d}(t),$$

where: $\vec{\rho}$: fixed vector from fulcrum to COM ; $\mathbf{R}(t)$: rotation around an oblique axis through $F(t)$
$\vec{d}(t)$: downward translation ($\dot{d}_z(t) < 0$)
Irreversible collapse occurs when:

$$\text{proj}_{\,B_U}(C_U(t)) \notin B_U$$

# APPENDIX C

## Illustrative Application of the Instability Framework: A Theoretical Walk-Through for Coaches

**Arbuzov vs. Grigalashvili — World Championship 2025**

This appendix provides a step-by-step theoretical interpretation of a real competition sequence using the Instability Framework and the Functional Instability Index $I(t)$. The numerical values are illustrative, not derived from measured kinematic data, and are intended to demonstrate how coaches may apply the model in practice.

## C.1 Frame-by-Frame Dynamical Interpretation

<table>
<tr>
<td>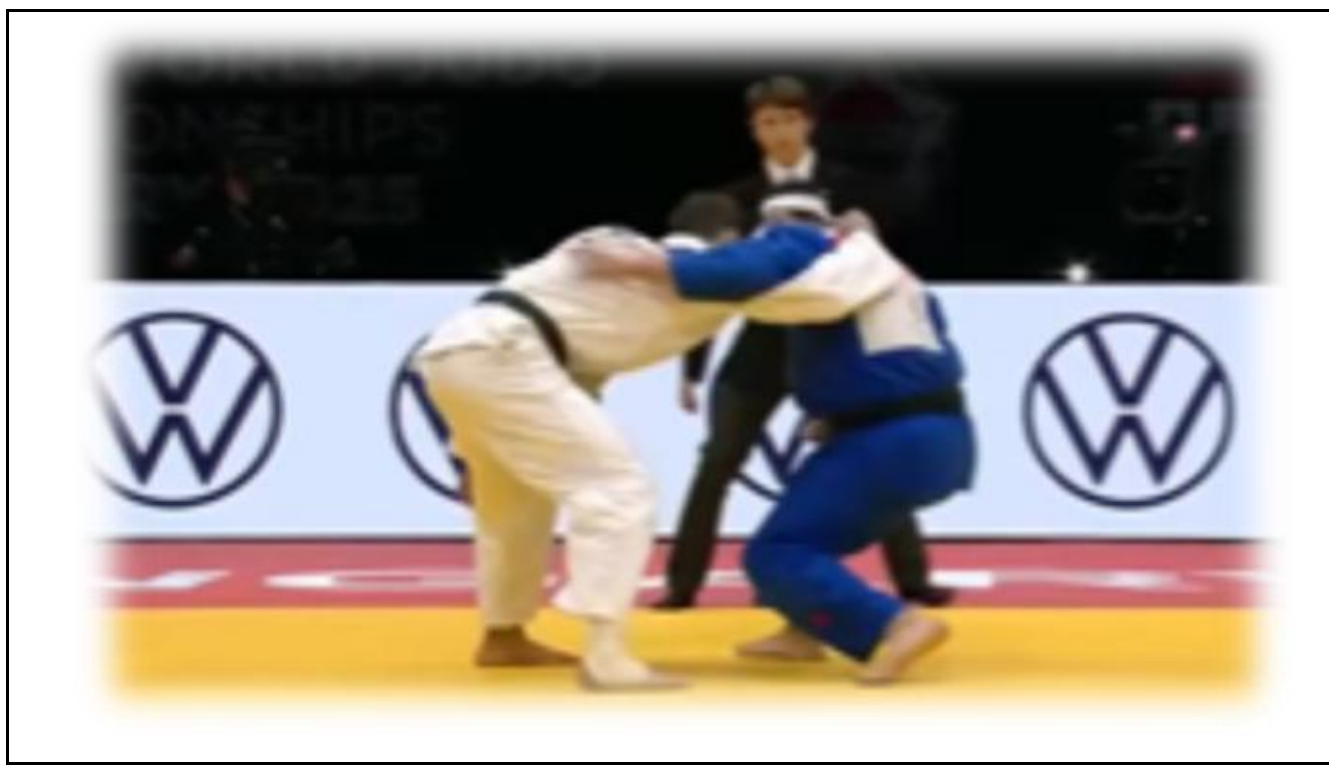</td>
<td>Frame 1 — 03:38 — Dyadic Equilibrium<br>State: Both athletes are within their stable basins of attraction.<br>Dynamics: Potential energy is balanced; transient static symmetry breaking is present. (Arbuzov).<br>Interpretation: Maximum stability; $I(t) = 0$.</td>
</tr>
</table>

| | |
|---|---|
| 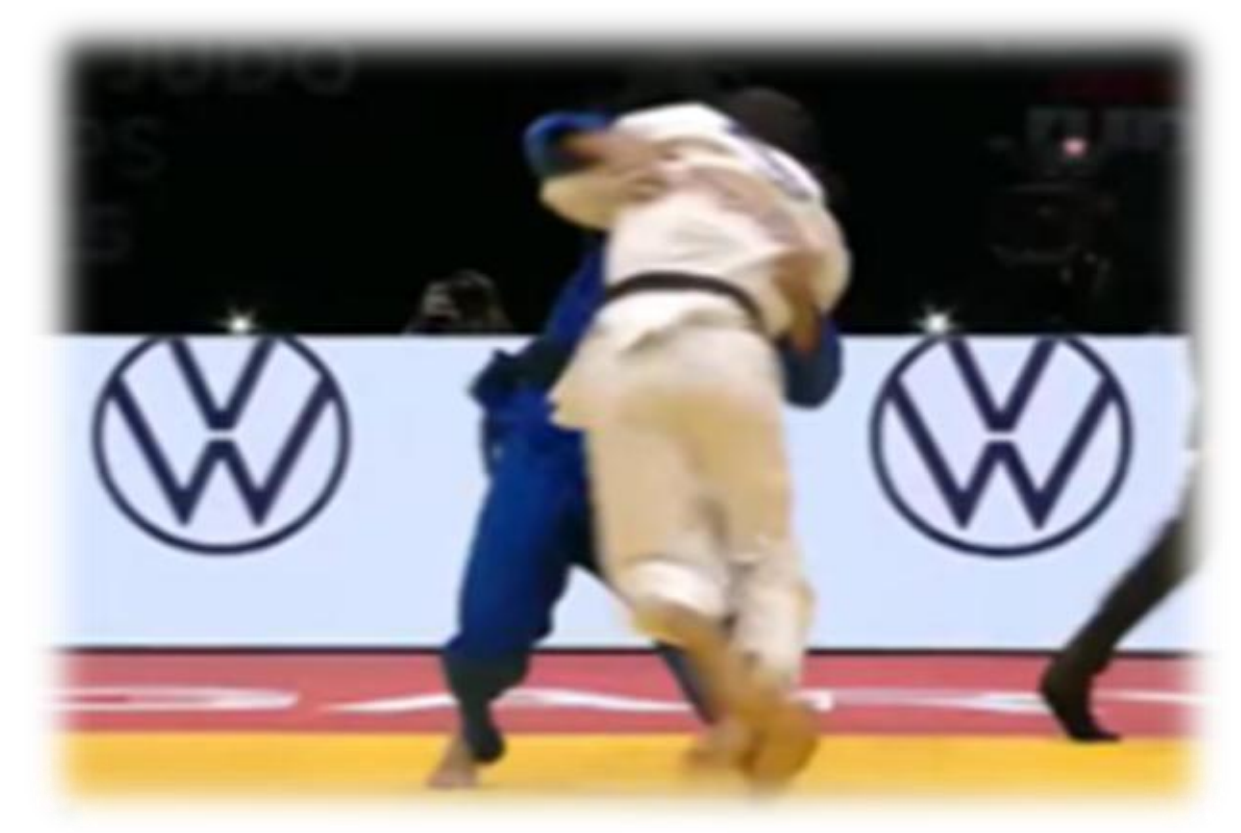 | ***Frame 2 — 03:39 — Dynamic Instability Trigger***<br>***Event:*** Grigalashvili (Blue) rightly, initiates dynamic instability against Arbuzov.<br>***Action***: He shifts rotating to increase directional stability in preparation for Yoko-sutemi, but positions himself slightly too far.<br>***Interpretation***: First dynamic symmetry-breaking event; instability begins to rise. |
| 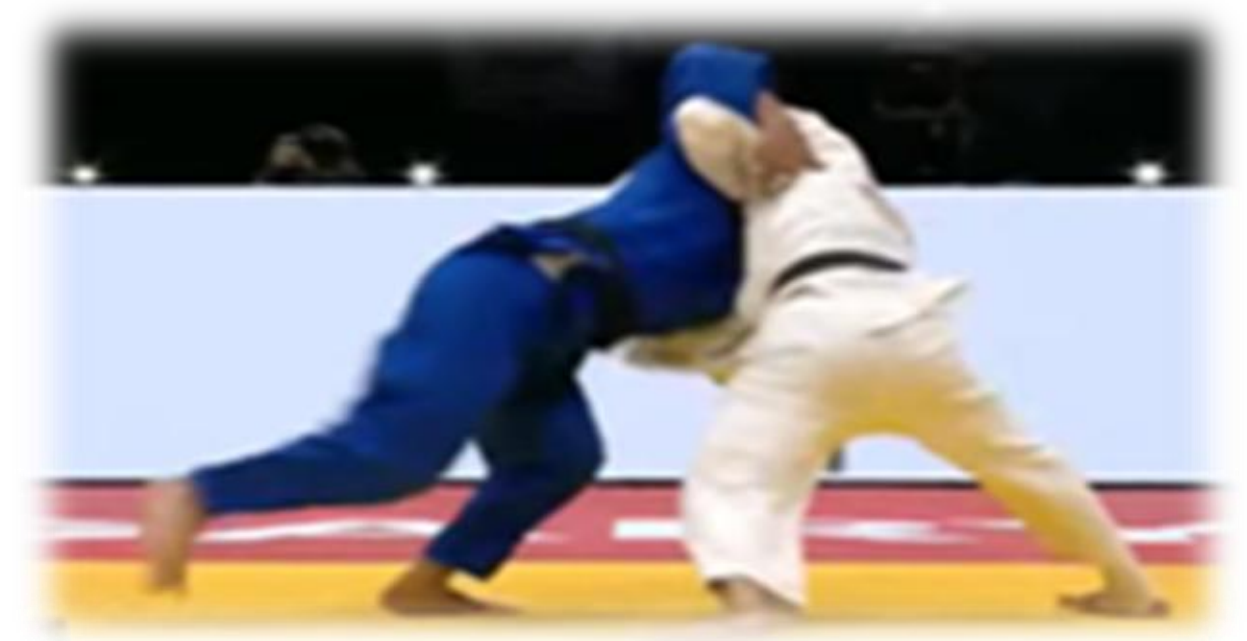 | ***Frame 3 — 03:40 A — The Bifurcation Point***<br>***Event***: Arbuzov (White) avoids the attack and intercepts the motion.<br>***State:*** The dyad reaches a saddle point, where the dyadic balance is critically compromised.<br>***Interpretation:*** This is the critical bifurcation; small perturbations determine the outcome. |
| 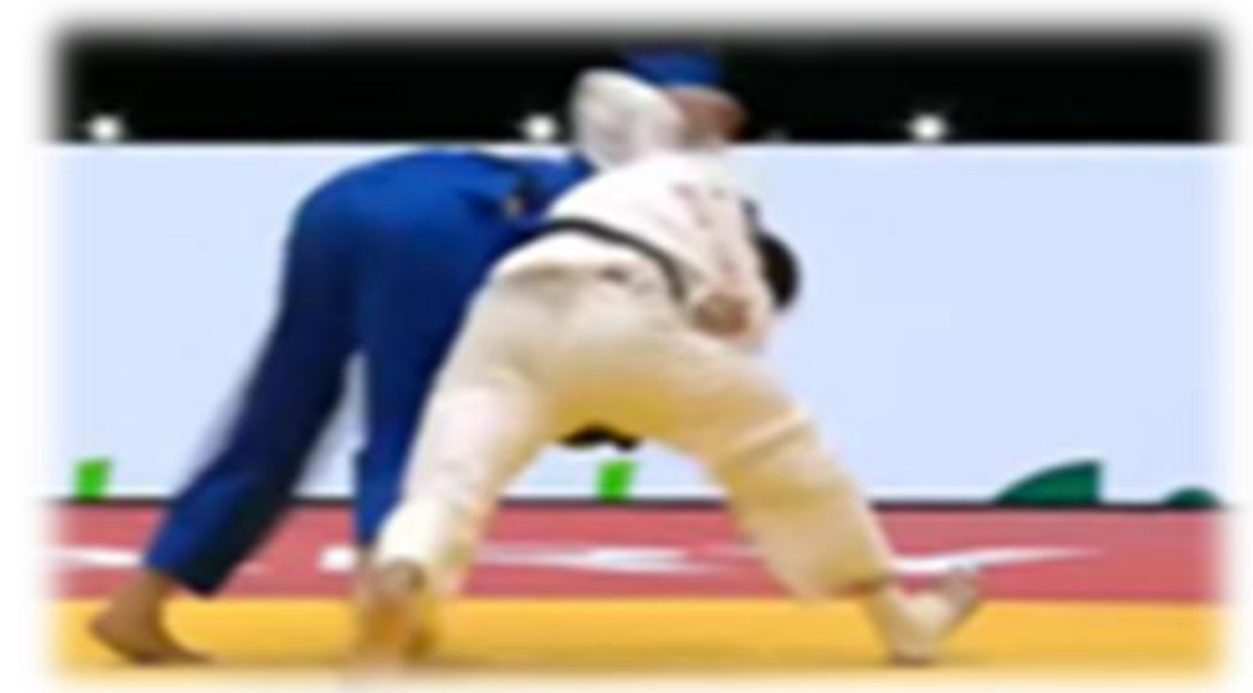 | ***Frame 4 — 03:40 B — Emerging FTLE Divergence***<br>***Event:*** The Finite-Time Lyapunov Exponent (FTLE) becomes positive.<br>***State***: The system enters a region of high sensitivity; trajectories diverge.<br>***Interpretation***: The dyad is now in the meta-stable Kuzushi window. |
| 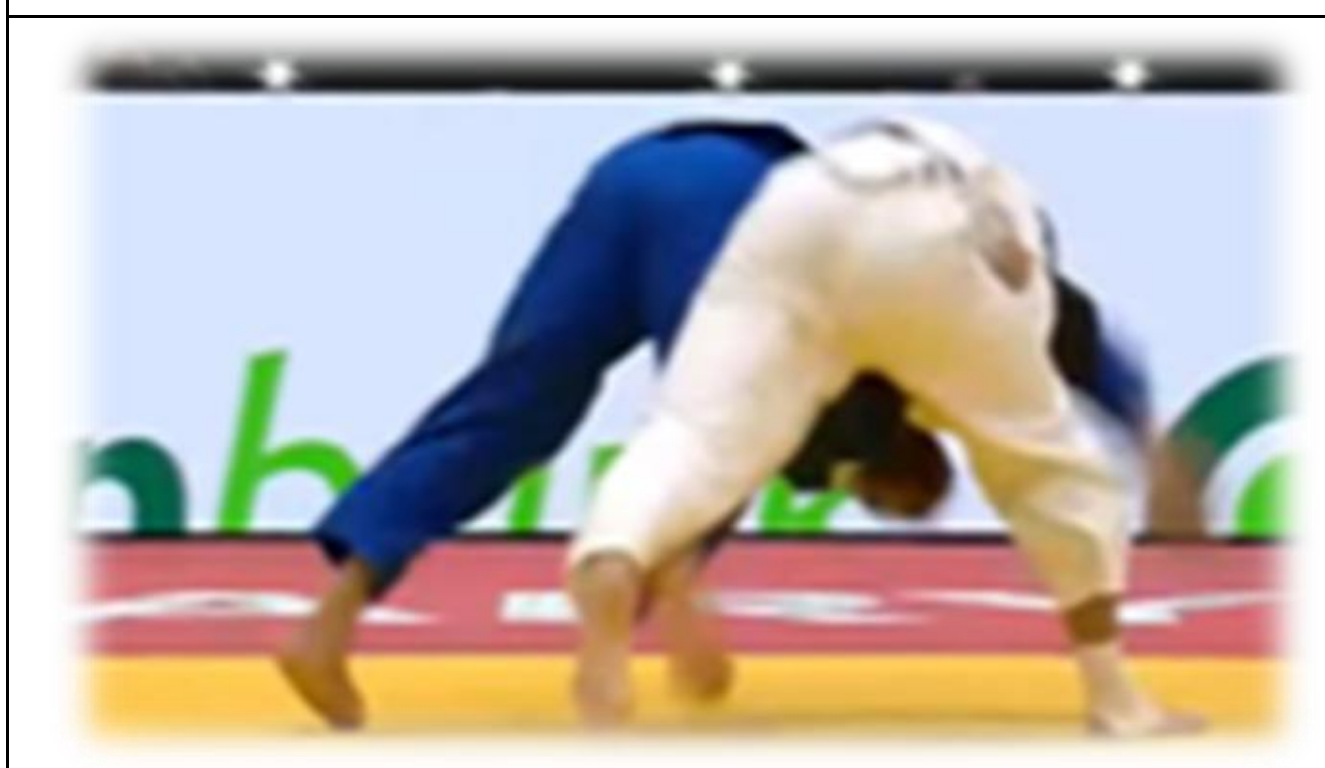 | ***Frame 5 — 03:40 C — Rotational Acceleration***<br>***Event***: Arbuzov redirects Grigalashvili's kinetic energy.<br>***State***: Rotational flow accelerates toward the floor.<br>***Interpretation***: Dynamic instability dominates; rotational attractor emerges. |
| 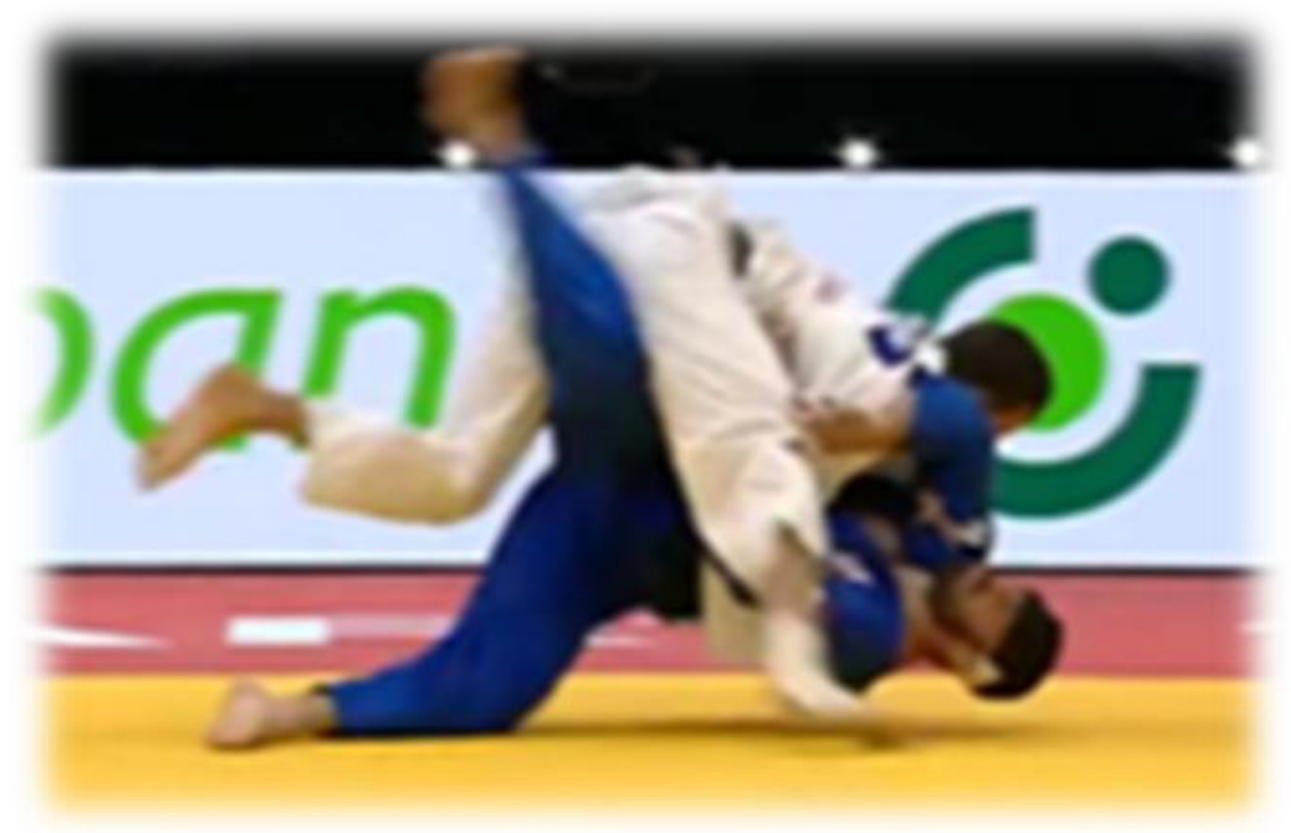 | ***Frame 6 — 03:40 D — Irreversible Collapse***<br>***System State***: The dyad transitions to a new attractor: the ground state (Ippon).<br>Grigalashvili is fully decoupled from the tatami, unable to return to his original stability basin of attraction.<br>Arbuzov maintains a stable dynamic axis while accelerating rotational velocity.<br>***Physical Concept:*** Total divergence; the evolution is now deterministic. |
| 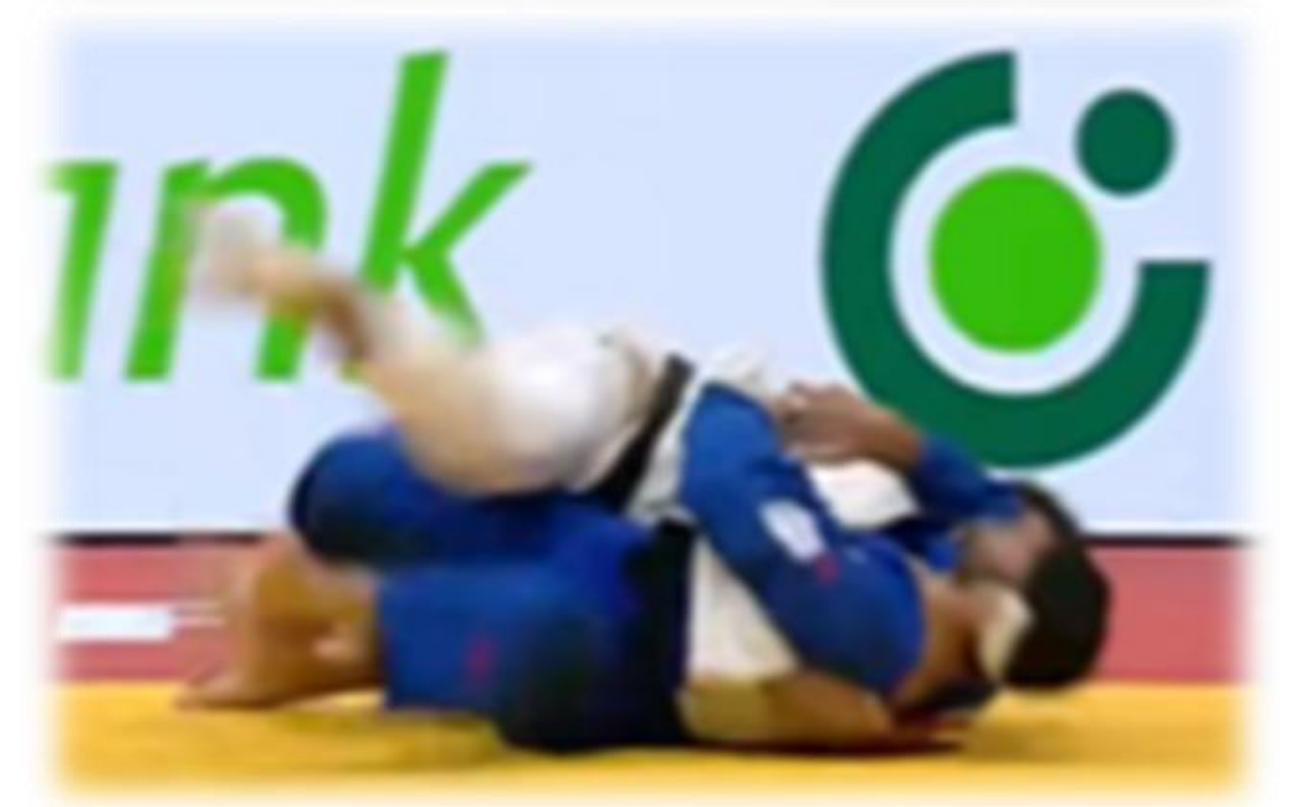 | ***Frame 7 — 03:40 E — System Decoupling***<br>***Event:*** Finalisation of the throw.<br>***State:*** Arbuzov re-establishes individual dynamic stability in Ne Waza; the dyad decouples. Outcome: Ippon. |

## C.2 Final Considerations for Coaches

Modern judo excellence is not merely the application of force, but the mastery of the dyadic phase space. A winning action occurs when Tori:

1. Recognises static instability in the opponent,
2. Transforms it into dynamic instability,
3. Maximises $\mu_{max} > 0$ and FTLE to ensure total divergence of Uke's trajectory,
4. Drives the system into irreversible gravitational collapse.

This is the essence of the perfect Ippon.

## C.3 Application of the Instability Index $I(t)$ A Theoretical Example for Coaches

The following example illustrates how the Functional Instability Index $I(t)$ can be applied to a real match sequence using the revised bifurcation-derived formula. $\chi(t) = \frac{\mu_{\max}(t)}{\eta(q(t))} \cdot \frac{K_0}{K_{\text{grip}}(t)+K_0}$ The three components μ_max(t), η(q(t)), and K_grip(t) are illustrated with plausible illustrative values.

The index is computed here in a **simplified didactic form** using the three analytical components of χ(t):

$I(t) = \frac{1}{1+e^{-k(\chi(t)-\chi_{\text{crit}})}}$ where:

- $\mu_{\max}(t)$ : maximum real eigenvalue of Jacobian A(t) — local dynamical divergence rate
- $\eta(q(t))$: normalised COM distance to support polygon boundary — geometric stability
- $K_{\text{grip}}(t)/K_0$: grip impedance ratio — defensive coupling modulation

## C.4 Table C

| Frame | Component Values (μ_max, η, Kgrip/$K_0$) | | Value | Sequential Evolution of $I(t)$ | |
|---|---|---|---|---|---|
| Time (mm:ss) | Illustrative values $\mu_{max}$, η, $K_{grip}/K_0$ | χ(t) = $\mu_{max}/\eta \cdot K_0/(K_{grip}+K_0)$ | $I(t)$ | Technical Phase | Dyadic State |
| 03:38 | (0.4, 0.3, 0.3) | (0.00, 0.00, 0.00) | 0.00 | Initial Equilibrium | Maximum stability |
| 03:39 | (0.3, 0.5, 0.2) | (0.20, 0.45, 0.07) | 0.30 | Grigalashvili Attack | Initiation of instability |
| 03:40 A | (0.2, 0.3, 0.5) | (0.45, 0.58, 0.87) | 0.70 | Saddle Point | Critical bifurcation |
| 03:40 E | (0.0, 1.0, 0.0) | (0.00, 1.00, 0.00) | 1.00 | Ippon | Total system saturation |

## C.5 Technical Note on Component Evolution

The shift in μ_max, η, and K_grip over 0.02 seconds is crucial — a timescale only AI can reliably detect.

- At equilibrium (03:38), $\mu_{\text{max}} \approx 0.4$η is near 1 (COM well inside support polygon): geometric stability dominates.
- During attack (03:39), $\eta(q) \approx 0.5\mu_{max}$ rises above 0: Jacobian eigenvalue divergence drives χ upward.
- At the saddle point (03:40 A), $\frac{K_{\text{grip}}}{K_0} \approx 0.5K_{grip}/K_0$ drops (grip weakened): $\chi_{\text{spikes}}$ — tactical timing
- De-ai moment of interception (De-ai) becomes decisive.

This demonstrates mathematically that Arbuzov's success is due to tactical synchronisation, not superior force. Note that the component values used here are purely illustrative and serve a didactic purpose; the complete revised $I(t)$ defined in Eq. (2) uses calibrated $\chi_{crit}$ and k parameters after supervised learning on labelled competition sequences.

## C.6 Conclusions

This dyadic model confirms that elite judo performance depends on the precise management of instability variables. The Arbuzov–Grigalashvili exchange illustrates how: tactical synchronisation can dominate kinetic initiation, instability can be amplified through timing rather than force, the Instability Index $I(t)$ can serve as a future metric for: coaching evaluation, athlete profiling, and potentially even Elo-style ranking based on the ability to generate instability.

# APPENDIX D

## A Three-Level Teaching Framework Based on Dynamic Instability

This appendix presents a structured pedagogical model derived from the nonlinear dynamical framework developed in the main text. Its purpose is to translate theoretical constructs—instability, symmetry breaking, attractors, FTLE, coupling stiffness, and the two archetypes—into a coherent, measurable, and progressive teaching methodology for coaches and athletes.

The model is articulated into three levels, each corresponding to a specific set of perceptual, coordinative, and tactical skills. Each level is associated with quantifiable indicators derived from the Functional Instability Index $I(t)$, symmetry metrics, COM displacement, and coupling dynamics.

### Level 1 – Perception of Instability and Static Asymmetry

*Objective:* Develop the athlete's ability to perceive the opponent's stability landscape, identifying static asymmetries, COM deviations, and grip-induced constraints before initiating dynamic actions.
*Theoretical Foundations:* Static symmetry breaking (postural asymmetry); COM offset $d_N(t)$; grip stiffness $K_g(t)$ as a stabilising parameter; Hurst parameter as an indicator of exploratory vs. deterministic behaviour.
*Key Skills:* Reading the opponent's stability gradient; identifying locked degrees of freedom; recognising exploitable postural biases; distinguishing stable vs. meta-stable configurations.
*Suggested Drills:* COM reading drills (identify lateral/sagittal COM shifts visually and via tactile feedback); head inclination; grip constraint mapping (classify grips by stabilising/destabilising effect); static asymmetry scenarios (Uke presents predefined asymmetries; Tori identifies optimal attack direction).
*AI Integration:* Automatic detection of COM deviation from the base of support; heat maps of static asymmetry; quantification of grip-induced stiffness.

### Level 2 – Manipulation of Instability and Dynamic Symmetry Breaking

*Objective:* Train the athlete to induce instability through controlled perturbations, rhythm changes, and dynamic coupling manipulation.
*Theoretical Foundations:* Dynamic symmetry breaking; local divergence (FTLE); coupling modulation through Kumi-kata; transitions between basins of attraction.
*Key Skills:* Creating micro-perturbations that accumulate into instability; manipulating rhythm to enter meta-stable regions; saturating Uke's degrees of freedom (virtual constraints) ex: wrapping grips; recognising the onset of positive FTLE.
*Suggested Drills:* Rhythm perturbation drills (alternating slow-fast entries); dynamic grip modulation (varying grip tension to alter coupling stiffness); meta-stability navigation (maintain the system near instability without crossing collapse threshold).
*AI Integration:* FTLE estimation over short windows; detection of timing mismatches; visualisation of coupling stiffness variations; identification of entry into meta-stable regions.

### Level 3 – Exploitation of Instability and Finalisation (Archetype-Specific)

*Objective:* Enable the athlete to finalise instability by guiding the dyad into the appropriate attractor rotational (Uchi-mata archetype) or gravitational collapse (Seoi-otoshi archetype).
*Theoretical Foundations:* Rotational attractor formation; gravitational discontinuity (catastrophic jump); downward acceleration $a_z(t)$; rotational torque $T_{\mathrm{rot}}(t)$; Instability Index $I(t)$ approaching 1.
*Key Skills:* Recognising the point of no return; executing the mechanical tool (lever or couple) at the singularity; maintaining flow continuity in rotational archetypes; maximising descent velocity in collapse archetypes.
*Suggested Drills:* For Uchi-mata (rotational instability): rotational flow continuity (continuous circular contact drills); hip–shoulder alignment timing synchronisation. For Seoi-otoshi (gravitational collapse): drop timing drills (practising drops faster than Uke's neuromuscular reaction time); COM vacuum creation (exercises focused on removing vertical support under Uke's COM).

*AI Integration:* Detection of rotational flow interruptions; measurement of descent velocity; identification of instability spikes; archetype classification through attractor shape.

**Summary of the Teaching Framework**

This three-level model provides a structured pathway from perception to manipulation to exploitation of instability. It aligns with the nonlinear dynamics of the Tori–Uke dyad and offers measurable indicators for each stage. The integration of AI tools enhances feedback precision and supports both athlete development and coaching strategies.